\documentclass[twocolumn,aps,superscriptaddress,showpacs,floatfix,prc]{revtex4}
\usepackage{mathrsfs}
\usepackage{amssymb}
\usepackage{amsmath}
\usepackage{graphicx}
\usepackage[normalem]{ulem}
\usepackage[dvips]{color}
\usepackage{bm}
\usepackage{longtable}

\renewcommand\sout{\bgroup \color{red} \ULdepth=-.5ex \ULset}

\renewcommand{\v}[1]{\textbf{#1}}
\renewcommand{\rm}[1]{\textrm{#1}}
\renewcommand{\d}{\mathrm{d}}

\begin{document}
\title{Symmetry energy of cold nucleonic matter within a relativistic mean field model encapsulating effects of high momentum nucleons induced by short-range correlations}

\author{Bao-Jun Cai\footnote{Email:landau1908feynman1918@gmail.com}}
\affiliation{Department of Physics and Astronomy, Texas A$\&$M
University-Commerce, Commerce, TX 75429-3011, USA}
\author{Bao-An Li\footnote{Corresponding author: Bao-An.Li$@$tamuc.edu}}
\affiliation{Department of Physics and Astronomy, Texas A$\&$M
University-Commerce, Commerce, TX 75429-3011, USA}
\date{\today}

\begin{abstract}
It is well known that short-range nucleon-nucleon correlations (SRC)
due to the tensor components and/or the repulsive core of nuclear
forces lead to a high (low) momentum tail (depletion) in the
single-nucleon momentum distribution above (below) the nucleon Fermi
surface in cold nucleonic matter. Significant progress has been made
recently in constraining the isospin-dependent parameters
characterizing the SRC-modified single-nucleon momentum distribution
in neutron-rich nucleonic matter using both experimental data and
microscopic model calculations. Using the constrained single-nucleon
momentum distribution in a nonlinear relativistic mean field (RMF)
model, we study the equation of state (EOS) of asymmetric nucleonic
matter (ANM), especially the density dependence of nuclear symmetry
energy $E_{\rm{sym}}(\rho)$. Firstly, as a test of the model, the
average nucleon kinetic energy extracted recently from
electron-nucleus scattering experiments using a neutron-proton
dominance model is well reproduced by the RMF model incorporating
effects of the SRC-induced high momentum nucleons, while it is
significantly under predicted by the RMF model using a step function
for the single-nucleon momentum distribution as in free Fermi gas
(FFG) models. Secondly, consistent with earlier findings within
non-relativistic models, the kinetic symmetry energy of
quasi-nucleons is found to be
$E^{\rm{kin}}_{\rm{sym}}(\rho_0)=-16.94\pm13.66\,\rm{MeV}$ which is
dramatically different from the prediction of
$E^{\rm{kin}}_{\rm{sym}}(\rho_0)\approx 12.5$ MeV by FFG models at
nuclear matter saturation density $\rho_0=0.16\,\rm{fm}^{-3}$.
Thirdly, comparing the RMF calculations with and without the high
momentum nucleons using two sets of model parameters both
reproducing identically all empirically constraints on the EOS of
symmetric nuclear matter (SNM) and the symmetry energy of ANM at
$\rho_0$, the SRC-modified single-nucleon momentum distribution is
found to make the $E_{\rm{sym}}(\rho)$ more concave around $\rho_0$
by softening it significantly at both sub-saturation and
supra-saturation densities, leading to an isospin-dependent
incompressibility of ANM in better agreement with existing
experimental data. Fourthly, the maximum mass of neutron stars is enhanced
by the increased kinetic pressure from high-momentum nucleons at supra-saturation densities in SNM.
\end{abstract}

 \maketitle

\section{Introduction}
The density dependence of nuclear symmetry $E_{\rm{sym}}(\rho)$ is
currently the most uncertain part of the equation of state (EOS) of
isospin asymmetric nucleonic matter (ANM) especially at
supra-saturation densities\,\cite{EPJA}. Owing to its importance in
both nuclear
physics\,\cite{LiBA98,Dan02,Bar05,Ste05,Che07a,LCK08,Tsa12,Che14}
and astrophysics\,\cite{Gle00,Lat04,Lat12,Lat14}, much efforts have
been devoted in recent years to constraining the
$E_{\rm{sym}}(\rho)$ using data from both terrestrial experiments
and astrophysical observations\,\cite{EPJA}. While significant
progress has been made in experimentally constraining the
$E_{\rm{sym}}(\rho)$ around the saturation density $\rho_0$, much
more work is needed to better constrain the $E_{\rm{sym}}(\rho)$ at
both sub-saturation and supra-saturation densities. On the other
hand, essentially all existing nuclear interactions have been used
in various many-body theories to calculate the $E_{\rm{sym}}(\rho)$.
While all models are tuned to be consistent with available
constrains on the $E_{\rm{sym}}(\rho)$ around $\rho_0$,  their
predictions diverge broadly at supra-saturation densities. For
uniform nucleonic matter, extensive studies have been underway by
various groups to understand why the $E_{\rm{sym}}(\rho)$ is so
uncertain especially at high densities and how one can better
constrain it. The spin-isospin dependence of three-body forces and
the isospin dependence of short-range nucleon-nucleon correlations
(SRC) induced by the poorly known nuclear tensor forces and the
repulsive core have been identified in several studies to be among
the main causes of the uncertainties in the $E_{\rm{sym}}(\rho)$ at
supra-saturation densities, see, e.g., refs. \cite{CXu10,Lee11,
Lee14,Wang12}. While at very low densities where cluster formation
and pairing become important, the $E_{\rm{sym}}(\rho)$ behaves
rather differently from expectations based on mean-field
models\,\cite{Joe,Typel,Mar}. Moreover, for clustered matter where
correlations dominate and the Coulomb force is important, there is
no neutron-proton exchange symmetry, it is even a question whether
it is necessary and how to introduce the symmetry energy in
describing the EOS of clustered matter.

How to relate isovector interactions with experimental observables
sensitive to the $E_{\rm{sym}}(\rho)$ has been a longstanding
question\,\cite{EPJA}. A thorough understanding about the origins
and properties of each part of the $E_{\rm{sym}}(\rho)$ is useful
for making further progress in this field. Usually, the symmetry
energy $E_{\rm{sym}}(\rho)$ can be decomposed into a kinetic and a
potential part, i.e.,
$E_{\rm{sym}}(\rho)=E_{\rm{sym}}^{\rm{kin}}(\rho)+E_{\rm{sym}}^{\rm{pot}}(\rho)$.
We emphasize that such a decomposition should be understood as for
quasi-nucleons of certain effective masses and momentum
distributions which are both determined by nuclear interactions.
Namely, the kinetic symmetry energy of quasi-nucleons also depends
on the interaction. However, in many analyses of data especially
using phenomenological models one often assumes that the kinetic
symmetry energy is that predicted by the free Fermi gas (FFG) model
for nucleons with bare masses and step functions for their momentum
distributions. The potential part is often parameterized with its
parameters extracted from fitting the data within adopted models for
describing the physics in question.

It is well known that short-range nucleon-nucleon correlations due
to the tensor components and/or the repulsive core of nuclear forces
lead to a high (low) momentum tail (depletion) in the single-nucleon
momentum distribution above (below) the nucleon Fermi surface in
cold nucleonic matter, see, e.g., refs.
\cite{Bethe,Ant88,Arr12,Cio15} for comprehensive reviews. In recent
years, significant efforts  have been made, e.g., refs.
\cite{Wei15,Wei15a,Hen14,Hen15,Col15,Egi06}, to constrain the
isospin-dependent parameters characterizing the SRC-modified
single-nucleon momentum distribution in neutron-rich nucleonic
matter using both experimental data and microscopic model
calculations. For instance, it has been found from analyzing
electron-nucleus scattering data that the percentage of nucleons in
the high momentum tail (HMT) above the Fermi surface is as high as
about 25\% in symmetric nuclear matter (SNM) but decreases gradually
to about only 1\% in pure neutron matter (PNM)\,\cite{Hen14,Hen15}.
Thus, the SRC-modified quasi-nucleon momentum distribution is
significantly different from the step function for the FFG at zero
temperature. Because of the momentum-squared weighting in
calculating the average nucleon kinetic energy, the strong isospin
dependence of the HMT makes the kinetic symmetry energy dramatically
different from the FFG model prediction using a step function for
the nucleon momentum
distribution\,\cite{CXu11,CXu13,Vid11,Lov11,Car12,Rio14,Car14,Hen15b,Cai15}.
In particular, the kinetic symmetry energy is significantly reduced
to even negative values in some model studies. In essence, the
symmetry energy is the energy difference between PNM and SNM in the
parabolic approximation of the ANM EOS. The neutron-proton
interaction dominated SRC increases significantly the average energy
per nucleon in SNM but has little effect on that in PNM, thus
leading to a reduction of the kinetic symmetry energy. This
expectation has been confirmed so far only within non-relativistic
approaches. It would be interesting to study effects of the HMT on
both the kinetic and potential parts of the $E_{\rm{sym}}(\rho)$
within a relativistic model.

The knowledge on each individual term of the $E_{\rm{sym}}(\rho)$ is
useful in both nuclear physics and astrophysics. For instance, in
simulating heavy-ion reactions using transport models one needs as
an input the potential symmetry energy of quasi-nucleons. Its
magnitude is limited by the total symmetry energy at $\rho_0$ known
to be around $31.6\pm 2.66$ MeV\,\cite{LiBA13} and the kinetic
symmetry energy normally assumed to be that predicted by the FFG
model. Several recent studies have shown that using a SRC-reduced
kinetic symmetry energy in transport model simulations can lead to
significant effects on isovector observables of heavy-ion
collisions\,\cite{Hen15b,Li15,Yon15,Li15a}. Interestingly, it was
also found recently that the critical densities and effects of the
formation of different charge states of $\Delta(1232)$ resonances in
neutron stars depend on how the kinetic and potential parts of the
$E_{\rm{sym}}(\rho)$ individually evolve as functions of
density\,\cite{Cai15a,Dra14}. Namely, in determining the critical
formation densities for $\Delta(1232)$ resonances in neutron stars
using chemical equilibrium conditions, the kinetic and potential
parts of the nucleon symmetry energy play different
roles\,\cite{Cai15a}. Basically, the $\Delta(1232)$ resonances
obtain a potential symmetry energy due to the $\tau_3(\Delta)\cdot
\tau_3(\rm{N})$ term in their interactions with nucleons where the
$\tau_3(\Delta)$ and $\tau_3(\rm{N})$ are the third component of the
isospin of $\Delta$ resonances and nucleons. However, the population
of $\Delta$ resonances is so low especially near their production
thresholds that they do not built their own Fermi spheres and thus
they do not have a kinetic symmetry energy. Depending on the
relative strengths of the $\rm{NN}\rho$ and $\Delta\Delta\rho$
coupling constants $g_{\rho\textrm{N}}$ and $g_{\rho\Delta}$, the
potential symmetry energies of the $\Delta$ resonances and nucleons
may completely cancel out but the kinetic symmetry energy of
nucleons remains in the equations determining the critical formation
densities of the four charge states of $\Delta$
resonances\,\cite{Cai15a}.

In this work, within a nonlinear relativistic mean field (RMF) model
incorporating the SRC-modified single-nucleon momentum distribution
with its parameters determined by electron-nucleus scattering
experiments and calculations using state-of-the-art many-body
theories, we study the EOS of ANM especially the
$E_{\rm{sym}}(\rho)$. Several interesting effects are found. In
particular,  comparing the RMF calculations with and without the HMT
using two sets of model parameters both reproducing identically all
empirically constraints on the EOS of SNM and the symmetry energy of
ANM at $\rho_0$, the SRC-modified nucleon momentum distribution
leads to a negative kinetic symmetry energy and the total symmetry
energy is softened at both sub-saturation and supra-saturation
densities. Moreover, only with the SRC-modified nucleon momentum
distribution, the recently extracted average kinetic energy per
nucleon from electron-nucleus scattering experiments can be
reproduced, providing a strong support for the existence of HMT in
nuclei. Furthermore, the HMT also enhance the maximum mass of neutron stars
by increasing the kinetic pressure of SNM at supra-saturation densities.

The paper is organized as follows, in Section \ref{SecII}, the
SRC-modified single-nucleon momentum distribution with a HMT and the
basic equations of the nonlinear RMF model are outlined. In Section
\ref{SecIII}, we evaluate the kinetic symmetry energy with the
SRC-modified single nucleon momentum distribution. Effects on the
EOS of SNM and the validation of the HMT are presented in Section
\ref{SecIV}. In Section \ref{SecV}, effects of the HMT on the
nucleon scalar density and Dirac effective mass are studied. Then in
Section \ref{SecVI} we examine effects of the HMT on the density
dependence of the total symmetry energy. In Section \ref{SecVII},
the effects of the HMT on the EOS of neutron star matter as well as
the mass-radius relation of neutron stars will be explored. Finally,
we summarize in Section \ref{SecVIII}. Detailed derivations for the
analytical expressions of the kinetic symmetry energy,
incompressibility coefficient $K_0$ of SNM and the slope parameter
$L$ of the symmetry energy within the RMF with HMT are given in the
three Appendixes.

\section{A Relativistic Mean Field Model Incorporating the SRC-Modified Nucleon Momentum
Distribution}\label{SecII}
In this section, we first summarize the main features and give all parameters of the SRC-modified
single-nucleon momentum distribution. Then we discuss how the relevant formalisms of the nonlinear RMF model
are generalized by replacing the previously used step function for the nucleon momentum distribution with the SRC-modified one including a high-momentum tail.
We notice that tensor forces have no effect at the mean-field level. The SRC-modified momentum distribution
can not be obtained self-consistently within the RMF model itself.

\subsection{The SRC-Modified Nucleon Momentum Distribution Function}
Here we briefly describe the SRC-modified single-nucleon momentum
distribution function encapsulating a high momentum tail used in the
present work. More details can be found in ref. \cite{Cai15}. The single-nucleon momentum distribution
function in ANM has the following form,
\begin{equation}\label{MDGen}
n^J_{\v{k}}(\rho,\delta)=\left\{\begin{array}{ll}
\Delta_J+\beta_J{I}\left(\displaystyle{|\v{k}|}/{k_{\rm{F}}^J}\right),~~&0<|\v{k}|<k_{\rm{F}}^J,\\
&\\
\displaystyle{C}_J\left({k_{\rm{F}}^{J}}/{|\v{k}|}\right)^4,~~&k_{\rm{F}}^J<|\v{k}|<\phi_Jk_{\rm{F}}^J.
\end{array}\right.
\end{equation}
Here, $J=\rm{n,p}$ is the isospin index,
$k_{\rm{F}}^J=k_{\rm{F}}(1+\tau_3^J\delta)^{1/3}$ is the Fermi
momentum where $k_{\rm{F}}=(3\pi^2\rho/2)^{1/3}$ and
$\tau_3^{\rm{n}}=+1$, $\tau_3^{\rm{p}}=-1$. It is worth emphasizing
that the above form of nucleon momentum distribution function is
consistent with the well-known predictions of microscopic nuclear
many-body theories\,\cite{Bethe,Ant88,Arr12,Cio15} and the recent
experimental
findings\,\cite{Hen14,Hen15,Wei15,Wei15a,Col15}.\begin{figure}[h!]
\centering
  \includegraphics[width=6.5cm]{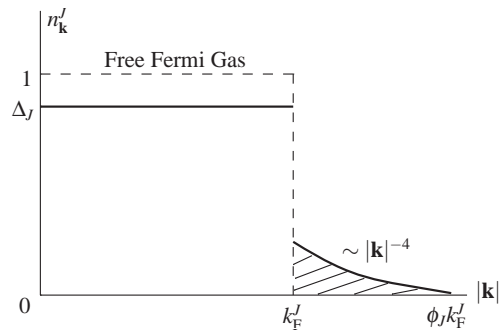}
  \caption{A sketch of the single-nucleon momentum distribution with a high momentum tail used in this work.
Taken from ref. \cite{Cai15}.}
  \label{mom-dis}
\end{figure}

In (\ref{MDGen}), the $\Delta_J$ measures the depletion of the Fermi
sphere at zero momentum with respect to the FFG model while
$\beta_J$ is the strength of the momentum dependence
$I(\v{k}/k_{\rm{F}}^J)$ of the depletion near the Fermi surface.
Owing to the small effects of $\beta_J$ on the energy per
nucleon\,\cite{Cai15}, we assume $\beta_J=0$ in this work. The
sketch of $n^J_{\v{k}}(\rho,\delta)$ is shown in Fig. \ref{mom-dis}.
The isospin structure of the parameters $\Delta_J$, $C_J$ and
$\phi_J$ is found to be
$Y_J=Y_0(1+Y_1\tau_3^J\delta)$\,\cite{Cai15}. The amplitude ${C}_J$
and high-momentum cutoff coefficient $\phi_J$ determine the fraction of nucleons
in the HMT via
\begin{equation}\label{xPNM}
x_J^{\rm{HMT}}=3C_{{J}}\left(1-\phi_J^{-1}\right).
\end{equation}
The normalization condition $
[{2}/{(2\pi)^3}]\int_0^{\infty}n^J_{\v{k}}(\rho,\delta)\d\v{k}=\rho_J={(k_{\rm{F}}^{J})^3}/{3\pi^2}
$ requires that only two of the three parameters, i.e., ${C}_J$,
$\phi_J$ and $\Delta_J$, are independent. Here we choose the first
two as independent and determine the $\Delta_J$ by
\begin{equation}\label{DeltaJ}
\Delta_J=1-3{C}_J(1-\phi_J^{-1}).
\end{equation}

The ${C}/{|\mathbf{k}|^4}$ shape of the HMT both for SNM and pure
neutron matter (PNM) is strongly supported by several recent studies
both theoretically and experimentally. Combining the results from
analyzing cross sections of $\rm{d}(\rm{e}, \rm{e}^{\prime}\rm{p})$
reactions\,\cite{Hen15} and medium-energy photonuclear
absorptions\,\cite{Wei15}, the $C_0$ was found to be
$C_0\approx0.161\pm0.015$. With this $C_0$ and the value of
$x^{\rm{HMT}}_{\rm{SNM}}=28\%\pm4\%$\,\cite{Hen14,Hen15,Hen15b}
obtained from systematic analyses of inclusive (e,e$'$) reactions
and data from exclusive two-nucleon knockout reactions, the HMT
cutoff parameter in SNM is determined to be
$\phi_0=(1-x_{\rm{SNM}}^{\rm{HMT}}/3{C}_0)^{-1}=2.38\pm0.56$\,\cite{Cai15}.
The value of $C_{\rm{n}}^{\rm{PNM}}=C_0(1+C_1)$ was extracted by
applying the adiabatic sweep theorem\,\cite{Tan08} to the EOS of PNM
constrained by predictions of microscopic nuclear many-body
theories\,\cite{Sch05,Epe09a,Tew13,Gez13,Gez10} and the EOS of cold
atoms under unitary condition\,\cite{Tan08,Stew10}. More
specifically,  $C_{\rm{n}}^{\rm{PNM}}\approx0.12$ and
$C_1=-0.25\pm0.07$ were obtained\,\cite{Cai15}. By inserting the
values of
$x_{\rm{PNM}}^{\rm{HMT}}=1.5\%\pm0.5\%$\,\cite{Hen14,Hen15,Hen15b}
extracted in the same way as the $x_{\rm{SNM}}^{\rm{HMT}}$ and
$C_{\rm{n}}^{\rm{PNM}}$ into Eq. ({\ref{xPNM}), the high momentum
cutoff parameter for PNM was determined to be
$\phi_{\rm{n}}^{\rm{PNM}}\equiv
\phi_0(1+\phi_1)=(1-x_{\rm{PNM}}^{\rm{HMT}}/3C_{\rm{n}}^{\rm{PNM}})^{-1}=1.04\pm0.02$\,\cite{Cai15}.
Consequently, $\phi_1=-0.56\pm0.10$\,\cite{Cai15} was obtained by
using the $\phi_0$ determined earlier.

\subsection{Basic Equations in the Nonlinear Relativistic Mean Field Model Incorporating the SRC-Modified Single Nucleon Momentum Distribution}
The nonlinear RMF model has been very successful in describing many nuclear
phenomena, see, e.g., refs. \cite{Ser86,Ser97,Rei89,Rin96,Men06}. In the following,
we outline major ingredients of the nonlinear RMF model we use in this work. The emphasis is on describing
where and how the SRC-modified nucleon momentum distribution is used to replace the FFG step function traditionally used in all RMF models.

The interacting Lagrangian of the nonlinear RMF model supplemented
with couplings between the isoscalar and the isovector mesons
reads\,\cite{Mul96,Hor01,Hor01a,Hor02,Tod05,Che07,Cai12,Fat10,Fat13,Dut14}
\begin{align}
\mathcal{L}=&\overline{\psi}\left[\gamma_{\mu}(i\partial^{\mu}-g_{\omega}\omega^{\mu}
-g_{\rho}\vec{\mkern1mu\rho}^{\mu}\cdot\vec{\mkern1mu\tau})-(M-g_{\sigma}\sigma)\right]\psi\notag\\
&-\frac{1}{2}m_{\sigma}^2\sigma^2+\frac{1}{2}\partial_{\mu}\sigma\partial^{\mu}\sigma
-U(\sigma)\notag\\
&+\frac{1}{2}m_{\omega}^2\omega_{\mu}\omega^{\mu}-\frac{1}{4}\omega_{\mu\nu}\omega^{\mu\nu}
+\frac{1}{4}c_{\omega}\left(g_{\omega}\omega_{\mu}\omega^{\mu}\right)^2\notag\\
&+\frac{1}{2}m_{\rho}^2\vec{\mkern1mu\rho}_{\mu}\cdot\vec{\mkern1mu\rho}^{\mu}
-\frac{1}{4}\vec{\mkern1mu\rho}_{\mu\nu}\cdot\vec{\mkern1mu\rho}^{\mu\nu}\notag\\
&+\frac{1}{2}g_{\rho}^2\vec{\mkern1mu\rho}_{\mu}\cdot\vec{\mkern1mu\rho}^{\mu}
\Lambda_{\mathrm{V}}g_{\omega}^2\omega_{\mu}\omega^{\mu},\label{NLRMF}
\end{align}
where $\omega_{\mu \nu }\equiv \partial _{\mu }\omega _{\nu
}-\partial _{\nu
}\omega _{\mu }$ and$~\vec{\mkern1mu\rho}_{\mu \nu }\equiv \partial _{\mu }\vec{%
\mkern1mu\rho }_{\nu }-\partial _{\nu }\vec{\mkern1mu\rho }_{\mu }$
are strength tensors for $\omega $ field and $\rho $ field,
respectively. $\psi $, $\sigma $, $\omega _{\mu }$,
$\vec{\mkern1mu\rho }_{\mu }$ are nucleon field, isoscalar-scalar
field, isoscalar-vector field and isovector-vector field,
respectively, and the arrows denote the vector in isospin space,
$U({\sigma})=b_{\sigma}M(g_{\sigma}\sigma)^3/3+c_{\sigma}(g_{\sigma}\sigma)^4/4
$ is the self interaction term for $\sigma$ field. $\Lambda
_{\textrm{V}}$ represents the coupling constant between the
isovector $\rho $ meson and the isoscalar $\omega $ meson. In
addition, $M=939\,\rm{MeV}$ is the nucleon mass and $m_{\sigma }$,
$m_{\omega }$, $m_{\rho }$ are masses of mesons.

In the mean field approximation, after neglecting effects of
fluctuation and correlation, meson fields are replaced by their
expectation values, i.e., $\overline{\sigma}\rightarrow \sigma $,
$\overline{\omega}_{0}\rightarrow \omega _{\mu } $,
$\overline{\rho}_{0}^{(3)}\rightarrow \vec{\mkern1mu\rho }_{\mu }$,
where subscript \textquotedblleft $0$" indicates zeroth component of
the four-vector, superscript \textquotedblleft ($3$)" indicates
third component of the isospin. Furthermore, we also use in this
work the non-sea approximation which neglects the effect due to
negative energy states in the
Dirac sea. The mean field equations are then expressed as 
\begin{align}
m_{\sigma }^{2}\overline{\sigma}=& g_{\sigma }\left[ \rho
_{\textrm{S}}-b_{\sigma }M\left(
g_{\sigma }\overline{\sigma}\right) ^{2}-c_{\sigma }\left( g_{\sigma }\overline{\sigma}%
\right) ^{3}\right], \label{eomf}\\
m_{\omega }^{2}\overline{\omega}_{0}=& g_{\omega }\left[ \rho
-c_{\omega }\left(
g_{\omega }\overline{\omega}_{0}\right) ^{3}-\Lambda _{\textrm{V}}g_{\omega }\overline{\omega}%
_{0}\left( g_{\rho }\overline{\rho}_{0}^{(3)}\right) ^{2}\right], \\
m_{\rho }^{2}\overline{\rho}_{0}^{(3)}=& g_{\rho }\left[ \rho
_{\textrm{p}}-\rho
_{\textrm{n}}-\Lambda _{\textrm{V}}g_{\rho }\overline{\rho}_{0}^{(3)}\left( g_{\omega }\overline{%
\omega}_{0}\right) ^{2}\right],
\end{align}%
where $ \rho =\langle \overline{\psi}\gamma ^{0}\psi \rangle =\rho
_{\textrm{n}}+\rho _{\textrm{p}}$ and $\rho _{\textrm{S}}=\langle
\overline{\psi}\psi \rangle =\rho _{\textrm{S,n}}+\rho
_{\textrm{S,p}}$ are the baryon density and scalar density,
respectively, with the latter given by
\begin{align}
\rho _{\textrm{S},J}=&\frac{2}{(2\pi)^3}\int_0^{\phi_Jk_{\rm{F}}^J}n_{\v{k}}^J\textrm{d}\textbf{k}\frac{%
M_J^{\ast }}{\sqrt{|\textbf{k}|^{2}+{M_J^{\ast }}^{2}}} \notag\\
=&\frac{2}{(2\pi
)^{3}}\int_{0}^{k_{\textrm{F}}^{J}}\Delta_J\textrm{d}\textbf{k}\frac{M_J^{\ast
}}{\sqrt{|\textbf{k}|^{2}+{M_J^{\ast
}}^{2}}}\notag\\
&+\frac{2}{(2\pi
)^{3}}\int_{k_{\rm{F}}^J}^{\phi_Jk_{\rm{F}}^J}C_J\left(\frac{k_{\rm{F}}}{|\v{k}|}\right)^4\d\v{k}
\frac{M_J^{\ast }}{\sqrt{|\textbf{k}|^{2}+{M_J^{\ast }}^{2}}}.
\label{ScaDen}
\end{align}
The change introduced by the SRC-modified nucleon momentum distribution is in the following replacement
\begin{equation}
\int_0^{k_{\rm{F}}^J}(\rm{FFG step function})f\d\v{k}\longrightarrow\int_0^{\phi_Jk_{\rm{F}}^J}n_{\v{k}}^J\,(\rm{HMT})f\d\v{k}
\end{equation}
with $f$ being any quantity. In the following, we often use the
``HMT model" in this work as the abbreviation for the nonlinear RMF
model using the SRC-modified nucleon momentum distribution, while
the ``FFG model" refers to the original nonlinear RMF model using
the FFG step function for the single-nucleon momentum distribution. The Fermi energy of
nucleon $J$ is
$E_{\rm{F}}^{J\ast}=(k_{\textrm{F}}^{J,2}+M_J^{\ast,2})^{1/2}$ where
$M^{\ast}_J$ is the nucleon Dirac mass defined as
\begin{align}
M^{\ast}_J\equiv M-g_{\sigma }\overline{\sigma}.
\end{align}

The energy-momentum density tensor for the interacting Lagrangian
density in Eq. (\ref{NLRMF}) can be written as
\begin{align}
\mathcal{T}^{\mu \nu }=&\overline{\psi}i\gamma ^{\mu }\partial ^{\nu
}\psi +\partial ^{\mu }\sigma \partial ^{\nu }\sigma\notag\\
&-\omega^{\mu \eta }\partial ^{\nu
}\omega _{\eta }-\vec{\mkern1mu \rho }^{\mu \eta }\partial ^{\nu }\vec{\mkern%
1mu\rho }_{\eta }-\mathcal{L}g^{\mu \nu },  \label{EnMomTen}
\end{align}%
where $g_{\mu \nu }=(+,-,-,-)$ is the Minkowski metric. In the mean
field approximation, the mean value of time (zero) component of the
energy-momentum density tensor is the energy density of the nuclear
matter
system, i.e.,%
\begin{align}
\varepsilon =& \langle \mathcal{T}^{00}\rangle \notag\\
=&\varepsilon^{\mathrm{kin}%
}_{\textrm{n}}+\varepsilon^{\mathrm{kin}}_{\textrm{p}}+\frac{1}{2}\left[ m_{\sigma }^{2}\overline{%
\sigma}^{2}+m_{\omega }^{2}\overline{\omega}_{0}^{2}+m_{\rho }^{2}\left( \overline{\rho%
}_{0}^{(3)}\right) ^{2}\right]  \notag \\
& +\frac{1}{3}b_{\sigma }(g_{\sigma
}\overline{\sigma})^{3}+\frac{1}{4}c_{\sigma
}(g_{\sigma }\overline{\sigma})^{4}+\frac{3}{4}c_{\omega }(g_{\omega }\overline{\omega}%
_{0})^{4}\notag\\
& +\frac{3}{2}\left( g_{\rho }\overline{\rho}_{0}^{(3)}\right) ^{2}\Lambda _{\textrm{V}}(g_{\omega }\overline{\omega}%
_{0})^{2},
\end{align}%
where%
\begin{align}
\varepsilon^{\mathrm{kin}}_{J} =&\frac{2}{(2\pi )^{3}}\int_{0}^{k_{\textrm{F}}^{J}}\Delta_J\textrm{d}%
\textbf{k}\sqrt{|\textbf{k}|^{2}+{M_J^{\ast
2}}}\notag\\
&+\frac{2}{(2\pi )^{3}}\int_{k_{\rm{F}}^J}^{\phi_Jk_{\textrm{F}}^{J}}C_J\left(\frac{k_{\rm{F}}^J}{|\v{k}|}\right)^4\textrm{d}%
\textbf{k}\sqrt{|\textbf{k}|^{2}+{M_J^{\ast 2}}}\label{EnDenKin}
\end{align}%
is the kinetic part of the energy density. Similarly, the mean value
of space components of the energy-momentum density tensor
corresponds to the
pressure of the system, i.e.,%
\begin{align}
P=& \frac{1}{3}\sum_{j=1}^{3}\langle \mathcal{T}^{jj}\rangle\notag\\
 =&P_{\mathrm{kin}%
}^{\textrm{n}}+P_{\mathrm{kin}}^{\textrm{p}}
 -\frac{1}{2}\left[ m_{\sigma }^{2}\overline{\sigma}^{2}-m_{\omega }^{2}\overline{%
\omega}_{0}^{2}-m_{\rho }^{2}\left( \overline{\rho}_{0}^{(3)}\right)
^{2}\right]
\notag \\
& -\frac{1}{3}b_{\sigma }(g_{\sigma
}\overline{\sigma})^{3}-\frac{1}{4}c_{\sigma
}(g_{\sigma }\overline{\sigma})^{4}+\frac{1}{4}c_{\omega }(g_{\omega }\overline{\omega}%
_{0})^{4}\notag\\
&+\frac{1}{2}\left( g_{\rho }\overline{\rho}_{0}^{(3)}\right) ^{2}\Lambda _{\textrm{V}}(g_{\omega }\overline{\omega}%
_{0})^{2} ,
\end{align}%
where the kinetic part of pressure is given by
\begin{align}
P_{\mathrm{kin}}^{J}=&\frac{1}{3\pi ^{2}}\int_{0}^{k_{\textrm{F}}^{J}}\Delta_J\textrm{d}k\frac{k^{4}}{%
\sqrt{k^{2}+{M_J^{\ast}}^{2}}}\notag\\
&+\frac{1}{3\pi ^{2}}\int_{k_{\rm{F}}^J}^{\phi_Jk_{\textrm{F}}^{J}}C_J\left(\frac{k_{\rm{F}}^J}{k}\right)^4\textrm{d}k\frac{k^{4}}{%
\sqrt{k^{2}+{M_J^{\ast}}^{2}}}. \label{pressureKin}
\end{align}

For completeness, in the following we recall the definitions of several physics quantities characterizing the EOS of SNM and the density dependence of nuclear symmetry energy around $\rho_0$.
Expressions of these quantities in the presence of the HMT are given in the appendixes. These expressions can be used readily to fix the RMF model parameters by reproducing the empirical values of these quantities at $\rho_0$. First of all, the EOS of ANM can be calculated through the energy density
$\varepsilon(\rho,\delta)$ by
\begin{equation}
E(\rho ,\delta )=\frac{\varepsilon (\rho ,\delta )}{\rho }-M.
\label{SiEn}
\end{equation}
One important relation holds between the pressure and the energy
density,
\begin{equation}
P=\rho^2\frac{\partial(\varepsilon(\rho,\delta)/\rho)}{\partial\rho},
\end{equation}
and in Appendix \ref{App3}, we will prove this relation in SNM.

The function $E(\rho, \delta)$ can be expanded as a power series of
even-order terms in $\delta $ as
\begin{equation}
E(\rho ,\delta )\simeq E_{0}(\rho )+E_{\text{sym}}(\rho )\delta ^{2}+%
\mathcal{O}(\delta ^{4}), \label{EoSpert}
\end{equation}%
where $E_{0}(\rho )=E(\rho ,\delta =0)$ is the EOS of SNM, and the
symmetry energy is expressed as
\begin{equation}
E_{\text{sym}}(\rho )= \left. \frac{1}{2}\frac{\partial ^{2}E(\rho
,\delta )}{\partial \delta ^{2}}\right\vert _{\delta =0}.
\label{DefEsym}
\end{equation}%
Around the saturation density $\rho _{0}$, the $E_{0}(\rho )$ can be
expanded, e.g., up to 2nd-order in density, as,
\begin{equation}
E_{0}(\rho )=E_{0}(\rho _{0})+\frac{1}{2}K_0\chi
^{2}+\mathcal{O}(\chi ^{3}), \label{DenExp0}
\end{equation}%
where $\chi =(\rho -\rho _{0})/3\rho _{0} $ is a dimensionless
variable characterizing the deviations of the density from the
saturation density $\rho _{0}$. The first term $E_{0}(\rho _{0})$ on
the right-hand-side of Eq. (\ref{DenExp0}) is the binding energy per
nucleon in SNM at $\rho _{0}$ and,
\begin{align}
K_{0} =&\left. 9\rho _{0}^{2}\frac{\text{d} ^{2}E_{0}(\rho
)}{\text{d} \rho ^{2}}\right\vert _{\rho =\rho _{0}}~~  \label{K0}
\end{align}%
is the incompressibility coefficient of SNM. Similarly, one can
expand the $E_{\mathrm{sym}}(\rho )$ around the normal density as
\begin{equation}
E_{\text{sym}}(\rho)=E_{\text{sym}}(\rho_0)+L\chi+\mathcal{O}(\chi^{2}),
\label{EsymLKr}
\end{equation}
with the slope parameter $L$ of the symmetry energy defined by
\begin{align}
L\equiv&\left.3\rho_0\frac{\text{d}E_{\mathrm{sym}}(\rho)}{\text{d}\rho}\right|_{\rho
=\rho_{0} }.
\end{align}

It is necessary to point out here an inconsistency of our approach. Since the phenomenological $n_{\v{k}}^J$ in Eq. (\ref{MDGen}) has no direct relation to the interacting
Lagrangian (\ref{NLRMF}), although it is constrained by recent experimental and microscopic theoretical studies, our results may have
some deviations from those using the $n_{\v{k}}^J$ obtained in models going
beyond the mean field approximation by solving the equation of $\psi$ in the presence of
interactions between nucleons and mesons expressed in the Lagrangian
(\ref{NLRMF}) self-consistently. In fact, this inconsistency exists
in almost all phenomenonlogical mean-field models. Ideally, one should first
reproduce quantitatively the experimentally constrained $n_{\v{k}}^J$ by adjusting parameters in the model Lagrangian.
Unfortunately, as shown by the strong model dependence in predicting the $n_{\v{k}}^J$ using various models and interactions,
our poor knowledge on the isospin dependence of short-range nucleon-nucleon interactions, such as the couplings $g_{\rho}$ and
$\Lambda_{\rm{V}}$ in (\ref{NLRMF}), still hinders reproducing quantitatively the experimentally constrained $n_{\v{k}}^J$.
Thus, our hybrid approach using directly the phenomenological $n_{\v{k}}^J$ constrained by experiments can give us some useful perspectives on
the effects of the HMT on the EOS of ANM.

\section{The Kinetic Symmetry Energy with High Momentum Nucleons}\label{SecIII}
As shown in detail in Appendix \ref{App1}, the kinetic symmetry
energy $E_{\rm{sym}}^{\rm{kin}}(\rho)$ in the RMF model with the HMT
can be written as
\begin{widetext}
\begin{align}
E_{\rm{sym}}^{\rm{kin}}(\rm{HMT})=&\frac{k_{\rm{F}}^2}{6E_{\rm{F}}^{\ast}}\left[1-3C_0\left(1-\frac{1}{\phi_0}\right)\right]
-3E_{\rm{F}}^{\ast}C_0\left[C_1\left(1-\frac{1}{\phi_0}\right)+\frac{\phi_1}{\phi_0}\right]\notag\\
&-\frac{9M_0^{\ast,4}}{8k_{\rm{F}}^3}\frac{C_0\phi_1(C_1-\phi_1)}{\phi_0}
\left[\frac{2k_{\rm{F}}}{M_0^{\ast}}\left(\left(\frac{k_{\rm{F}}}{M_0^{\ast}}\right)^2+1\right)^{3/2}\right.
\left.-\frac{k_{\rm{F}}}{M_0^{\ast}}\left(\left(\frac{k_{\rm{F}}}{M_0^{\ast}}\right)^2+1\right)^{1/2}-\rm{arcsinh}\left(\frac{k_{\rm{F}}}{M_0^{\ast}}\right)\right]\notag\\
&+\frac{2k_{\rm{F}}C_0(6C_1+1)}{3}\left[\rm{arcsinh}\left(\frac{\phi_0k_{\rm{F}}}{M_0^{\ast}}\right)-
\sqrt{1+\left(\frac{M_0^{\ast}}{\phi_0k_{\rm{F}}}\right)^2}\right.
\left.-\rm{arcsinh}\left(\frac{k_{\rm{F}}}{M_0^{\ast}}\right)+
\sqrt{1+\left(\frac{M_0^{\ast}}{k_{\rm{F}}}\right)^2}\right]\notag\\
&+\frac{3k_{\rm{F}}C_0}{2}\Bigg[\frac{(1+3\phi_1)^2}{9}\left(\frac{\phi_0k_{\rm{F}}}{F_{\rm{F}}^{\ast}}
-\frac{2F_{\rm{F}}^{\ast}}{\phi_0k_{\rm{F}}}\right)
+\frac{2F_{\rm{F}}^{\ast}(3\phi_1-1)}{9\phi_0k_{\rm{F}}}
-\frac{1}{9}\frac{k_{\rm{F}}}{E_{\rm{F}}^{\ast}}+\frac{4E_{\rm{F}}^{\ast}}{9k_{\rm{F}}}\Bigg]\notag\\
&+\frac{C_0(4+3C_1)}{3}\left[\frac{F_{\rm{F}}^{\ast}(1+3\phi_1)}{\phi_0}-E_{\rm{F}}^{\ast}\right],\label{EsymkinHMT}
\end{align}\end{widetext}
where
\begin{equation}E_{\rm{F}}^{\ast}=(M_0^{\ast,2}+k_{\rm{F}}^2)^{1/2} \rm{and}~
F_{\rm{F}}^{\ast}=(M_0^{\ast,2}+(\phi_0k_{\rm{F}})^2)^{1/2}.\end{equation}
In the FFG limit, $\phi_0=1,\phi_1=0$, only the first term of the
above expression survives and leads to $
E_{\rm{sym}}^{\rm{kin}}(\rho)\to
E_{\rm{sym}}^{\rm{kin}}(\rm{FFG})\equiv
k_{\rm{F}}^2/6E_{\rm{F}}^{\ast} $ as in traditional RMF models. The
kinetic symmetry energy in the presence of HMT is a function only of
the Dirac effective mass $M_0^{\ast}$. Using the values of
$C_0,C_1,\phi_0$ and $\phi_1$ given in the last section, we show in
Fig. \ref{fig-EsymkinRMFHMTDirac} the kinetic symmetry energy as a
function of $M_0^{\ast}$ at $\rho_0=0.16\,\rm{fm}^{-3}$ for both the
FFG and HMT models. In the whole range of $M_0^{\ast}$ considered as
reasonable, the kinetic symmetry energy at $\rho_0$ in the HMT model
is always negative.
\begin{figure}[h!]
\centering
  \includegraphics[width=7.5cm]{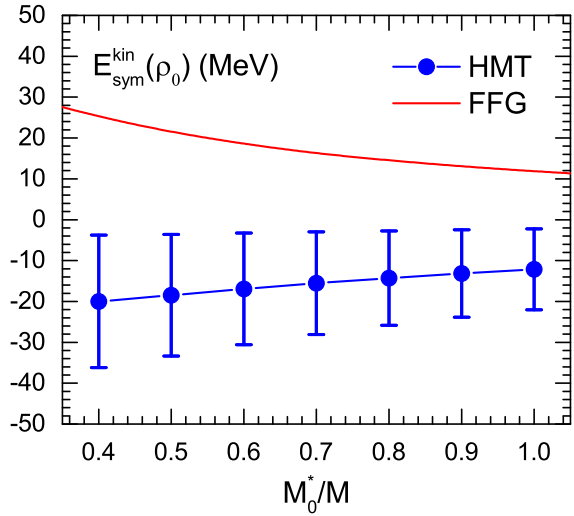}
  \caption{(Color Online) The kinetic symmetry energy as a function of Dirac effective mass of nucleon in SNM both in the FFG model and in the HMT model,
  $\rho_0=0.16\,\rm{fm}^{-3}$.}
  \label{fig-EsymkinRMFHMTDirac}
\end{figure}
For example, with $M_0^{\ast}/M=0.6$ the kinetic symmetry energy is
\begin{equation}
E_{\rm{sym}}^{\rm{kin}}(\rho_0)=-16.94\pm13.66\,\rm{MeV}
\end{equation}
where the errors are all from the uncertainties of $C_0,C_1,\phi_0$
and $\phi_1$. This value is close to the non-relativistic result of
$E_{\rm{sym}}^{\rm{kin}}(\rho_0)=-13.90\pm11.54\,\rm{MeV}$ according
to the expression\,\cite{Cai15}
\begin{align}
&E_{\rm{sym}}^{\rm{kin}}(\rho)=k_{\rm{F}}^2/6M\cdot[1+C_0(1+3C_1)(5\phi_0+3/\phi_0-8)\notag\\
+&3C_0\phi_1(1+3C_1/5)(5\phi_0-3/\phi_0) +27C_0\phi_1^2/5\phi_0].
\end{align}
Thus, the reduction of the kinetic symmetry energy in the presence
of HMT is general in both relativistic and non-relativistic
calculations\,\cite{Hen15b,Cai15}. Moreover, this result is also
consistent with the findings of several recent studies of the
kinetic EOS considering the SRC using both phenomenological models
and microscopic many-body
theories\,\cite{CXu11,CXu13,Vid11,Lov11,Car12,Rio14,Car14}.

\section{Validation of SRC-Modified Single-Nucleon Momentum Distribution and its Effects on the EOS of Symmetric Nuclear Matter}\label{SecIV}
First of all, it is necessary to point out that since we fixed the
parameters of the nucleon momentum distribution by using
experimental data and/or model calculations at the saturation
density, the possible density dependence of those parameters, i.e.,
$C_0,C_1,\phi_0$ and $\phi_1$ is not explored in this work as well
as in ref. \cite{Cai15}. The density dependence of the various terms
in the kinetic EOS is thus only due to that of the meson fields and
the Fermi momenta. In this section, all analytical expressions are
obtained under this assumption. The numerical results are obtained
by setting $\phi_0=2.38,\phi_1=-0.56,C_0=0.161$ and $C_1=-0.25$.  In
both the HMT and FFG models, the masses of meson fields are chosen
as $m_{\sigma}=500\,\mathrm{MeV}$, $m_{\omega}=782.5\,\mathrm{MeV}$,
and $m_{\rho}=763\,\mathrm{MeV}$.
\renewcommand*\tablename{\small Table}
\begin{table}[tbh]
\caption{Coupling constants used in the two RMF models  (right side)
and some empirical properties of asymmetric nucleonic matter used to
fix them (left side).}
\label{RMF_tab}%
\centering
\begin{tabular}{lr||lrr}
\hline\hline Quantity& this work & Coupling & FFG&HMT\\
\hline $\rho_0$ (fm$^{-3}$) & $0.15$ & $g_{\sigma}$ &
10.9310&10.8626\\
\hline $E_0(\rho_0)$
$(\rm{MeV})$&$-16.0$&$g_{\omega}$&14.5947&12.9185\\
\hline $M_0^{\ast}/M$ &0.6&$b_{\sigma}$&0.0007473&0.002119\\
\hline $K_0$ ($\rm{MeV}$) &230.0&$c_{\sigma}$&0.003882&$-0.0005139$\\
\hline $E_{\rm{sym}}(\rho_0)$ ($\rm{MeV}$)
&31.6&$g_{\rho}$&5.9163&7.8712\\
\hline $L$ ($\rm{MeV}$)&58.9&$\Lambda_{\rm{V}}$&0.2736&0.03740
\\
\hline\hline
\end{tabular}%
\end{table}

To determine the EOS and total symmetry energy with the SRC-modified
single-nucleon momentum distribution, we need to readjust the RMF
model parameters to reproduce all known empirical properties of SNM
and ANM. Thus, analytical expressions for quantities characterizing
these properties are necessary. Combining the results derived in
detail in the Appendixes, we have expressions for four such
quantities for SNM, i.e., the Dirac effective mass $M_0^{\ast}$; the
binding energy of SNM obtained through
$E_0(\rho)=\varepsilon_0(\rho)/\rho-M$ with $\varepsilon_0$ given by
(\ref{eps0rho}) together with (\ref{eps0rho1}) and (\ref{eps0rho2});
the pressure $P_0$ of SNM (\ref{p0rho}) and the incompressibility
coefficient $K_0$ of SNM (\ref{K0rho}). For ANM, we also need
expressions for the total symmetry energy $E_{\rm{sym}}(\rho)$ and
its slope $L$. While the kinetic symmetry energy
$E_{\rm{sym}}^{\rm{kin}}(\rho)$ is already given by
(\ref{EsymkinHMT}), the potential symmetry energy
$E_{\rm{sym}}^{\rm{pot}}(\rho)$ can be written
as\,\cite{Che07,Cai12}
\begin{equation}\label{Esympot}
E_{\rm{sym}}^{\rm{pot}}(\rho)=\frac{g_{\rho}^2\rho}{2Q_{\rho}}~{\textrm{with}}~
Q_{\rho}=m_{\rho}^2+\Lambda_{\rm{V}}g_{\rho}^2g_{\omega}^2\overline{\omega}_0^2.
\end{equation}
Correspondingly, the slope parameter $L$ of the total symmetry energy also has two parts, i.e., (\ref{TOTL}) with
the kinetic part $L^{\rm{kin}}$ given by (\ref{Lkin}) and the potential part $L^{\rm{pot}}$ by
(\ref{Lpot}). The total number of the analytical expressions is now six
while there are seven coupling constants in the Lagrangian density
(\ref{NLRMF}). We are thus still free to choose one of the seven coupling
constants, and in this work we fix the value of $c_{\omega}=0.01$
which is the same as that in the FSUGold parametrization\,\cite{Tod05}.
In this way, given the values of $M_0^{\ast}$, $E_0(\rho_0)$,
$\rho_0$, $K_0$, $E_{\rm{sym}}(\rho_0)$ and $L$, we can uniquely
determine the other six coupling constants. Listed in Table \ref{RMF_tab} are the coupling constants in both the FFG and HMT
models obtained from reproducing the same values of the listed empirical properties of ANM.
The value of $K_0=230\pm20\,\rm{MeV}$ was determined from
analyzing nuclear giant resonances (GMR)\,\cite{You99,Shl06,Pie10,Che12,Col14}.
For the $E_\text{sym}(\rho_0)$ and $L $, all existing constraints extracted so far
from both terrestrial laboratory measurements and astrophysical
observations are found to be essentially consistent with the 2013 global averages of
$E_{\text{sym}}({\rho _{0}}) = 31.6\pm2.66$ MeV and $L = 58.9 \pm
16$ MeV\,\cite{LiBA13}. We notice that the values of the two isovector parameters $g_{\rho}$
and $\Lambda_{\rm{V}}$ are significantly different in the HMT and FFG models.

\begin{figure}[h!]
\centering
  \includegraphics[width=8.8cm]{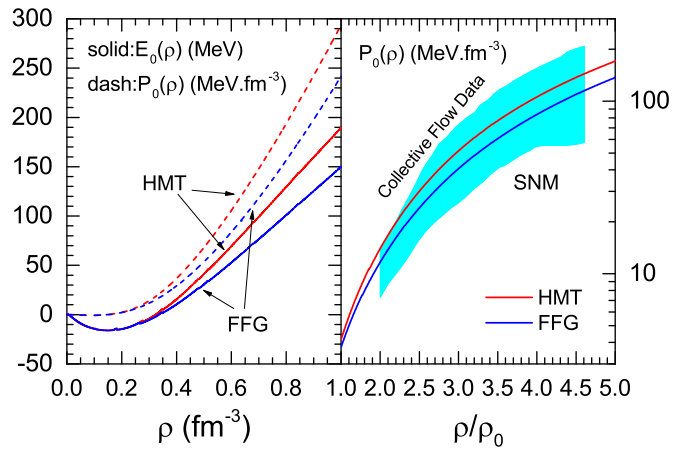}
  \caption{(Color Online) Left panel: The EOS and pressure $P_0$ of SNM as functions of density for both the FFG and HMT models;
  Right panel: a comparison between the model pressure $P_0$ of SNM with the experimental constraints from analyzing nuclear collective flows in heavy ion collisions.}
  \label{fig-E0RMFHMT}
\end{figure}
To evaluate the HMT and FFG models, we show in the left panel of
Fig. \ref{fig-E0RMFHMT} the binding energy and pressure of SNM as
functions of density. It is interesting to see that the HMT model
predicts a harder EOS for SNM at supra-saturation densities than the
FFG model while by design they both have the same values of
$M_0^{\ast}$, $\rho_0$, $E_0(\rho_0)$ and $K_0$. This is simply
because of the large contribution to the kinetic EOS by the high
momentum nucleons in the HMT model. Therefore, the HMT is expected
to affect the high order characteristic coefficients of the SNM at
$\rho_0$ compared to calculations with the FFG. More quantitatively,
the third-order Taylor expansion coefficient of the EOS of SNM
around $\rho_0$, i.e., the skewness of the SNM
$Q_0\equiv27\rho_0^3\partial^3E_0(\rho)/\partial\rho^3|_{\rho=\rho_0}$\,\cite{Che09,Che11,Cai14}
is changed from $Q_0^{\rm{FFG}}\approx-454\,\rm{MeV}$ in the FFG
model to $Q_0^{\rm{FFG}}\approx-266\,\rm{MeV}$ in the HMT model.
Unfortunately, our current knowledge on the parameter
$Q_0$\,\cite{Cai14,Far97,Ste10,Che11,Mei13,Sel14,San15} is still too
poor to put a constraint on it. On the other hand, the pressure of
SNM in the density range of about $2\rho_0$ to $5\rho_0$ has been
experimentally constrained by measuring nuclear collective flows in
heavy-ion collisions\,\cite{Dan02}, which is shown as a cyan band in
the right panel of Fig. \ref{fig-E0RMFHMT}. Although the HMT makes
skewness of the SNM higher, it is seen that the pressure of SNM in
the presence of HMT can still pass through the constraints from the
collective flow data. Namely, the uncertainty band of the
constraints on the EOS at supra-saturation densities is too broad to
distinguish the HMT and FFG predictions. Thus, as we shall discuss
next, additional experimental constraints are necessary to
distinguish the two models.

\begin{figure}[h!]
\centering
  \includegraphics[width=8.cm]{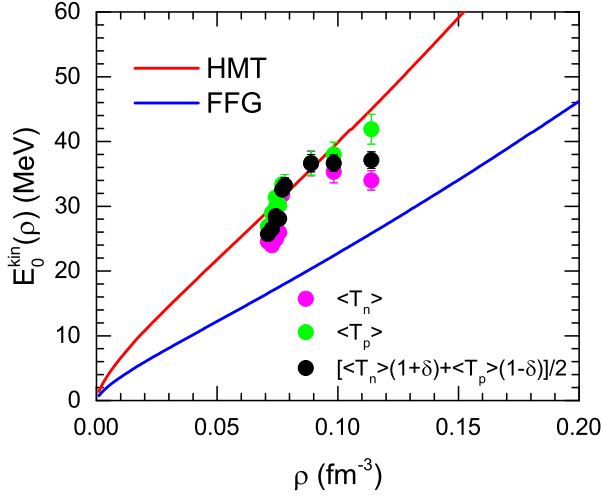}
  \caption{(Color Online) The kinetic EOS of SNM defined by (\ref{E0kin}).
The experimental kinetic energy of neutrons and protons in C, Al, Fe and Pb with error bars\,\cite{Hen14} and
$^{7,8,9}\textrm{Li}$, $^{9,10}\textrm{Be}$ and
$^{11}\textrm{B}$\,\cite{Sar14} were extracted using the neutron-proton dominance model.}
  \label{fig-Ekin0RMFHMT}
\end{figure}

In the nonlinear RMF model, the kinetic EOS of SNM is defined as
\begin{equation}
E_0^{\rm{kin}}(\rho)\equiv\frac{1}{\rho}\frac{2}{(2\pi)^3}\int_0^{\phi_0k_{\rm{F}}}n_{\v{k}}^0\sqrt{\v{k}^2+M_0^{\ast,2}}\d\v{k}-M_0^{\ast},\label{E0kin}
\end{equation}
where $n_{\v{k}}^0$ is the momentum distribution of nucleons in SNM.
The HMT and FFG model predictions for the $E_0^{\rm{kin}}(\rho)$ are
shown in Fig. \ref{fig-Ekin0RMFHMT}. Recently, the average kinetic
energy of neutrons and protons in C, Al, Fe and Pb with error bars
as well as $^{7,8,9}\textrm{Li}$, $^{9,10}\textrm{Be}$ and
$^{11}\textrm{B}$ without error bars were extracted from several
electron-nucleus scattering experiments using a neutron-proton
dominance model\,\cite{Hen14,Sar14}. We can translate the
$A$-dependence of the nucleon kinetic energy into its density
dependence through a well-established empirical
relationship\,\cite{Cen09,Che11a,Maz13,Mye69,Dan03,Dan09,Dan14}
\begin{equation}
\rho_{A}\simeq\frac{\rho_0}{1+\alpha/A^{1/3}}
\end{equation}
where $\alpha$ reflects the balance between the volume and surface
symmetry energies and in our calculation we adopt
$\alpha=2.8$\,\cite{Dan03} appropriate for the mass range
considered. The black points represent the average kinetic energy
per nucleon for these nuclei, i.e., $\langle T\rangle=[\langle
T_{\rm{n}}\rangle(1+\delta)+\langle T_{\rm{p}}\rangle(1-\delta)]/2$.
According to the parabolic approximation for the EOS of ANM, i.e.,
$E^{\rm{kin}}_{\rm{ANM}}(\rho)\simeq
E^{\rm{kin}}_{0}(\rho)+\delta^2E^{\rm{kin}}_{\rm{sym}}(\rho)$, even
for the most neutron-rich nucleus considered $^{208}\rm{Pb}$ with an
isospin asymmetry $\delta^2\simeq0.045$, we still have
$E^{\rm{kin}}_{\rm{ANM}}(\rho)\simeq E_{0}^{\rm{kin}}(\rho)$. This
means that the data in Fig. \ref{fig-Ekin0RMFHMT} are approximately
equal to the kinetic EOS of SNM $E_{0}^{\rm{kin}}(\rho)$. It is very
interesting to see that  the HMT prediction can well reproduce while
the FFG prediction falls about 40\% below the data around
$\rho_A=0.1$ fm$^{-3}$. This clearly indicates the importance of the
HMT in the SRC-modified single nucleon momentum distribution. It is
well known that mean-field models fail to describe the spectroscopic
factors extracted from electron scatterings on nuclei from $^{7}$Li
to $^{208}$Pb by about 30-40\% due to the lack of occupations of
energetic orbitals in these models where the short-range
correlations are not considered\,\cite{Lap93}. The observation here
that the FFG model under predicts the average nucleon kinetic energy
is due to the same reason and it misses the data by about the same
magnitude as in describing the spectroscopic factors.

\section{Nucleon Scalar Density and Dirac Effective Mass in the RMF Model with High Momentum Nucleons}\label{SecV}
\begin{figure}[h!]
\centering
  \includegraphics[width=8.5cm]{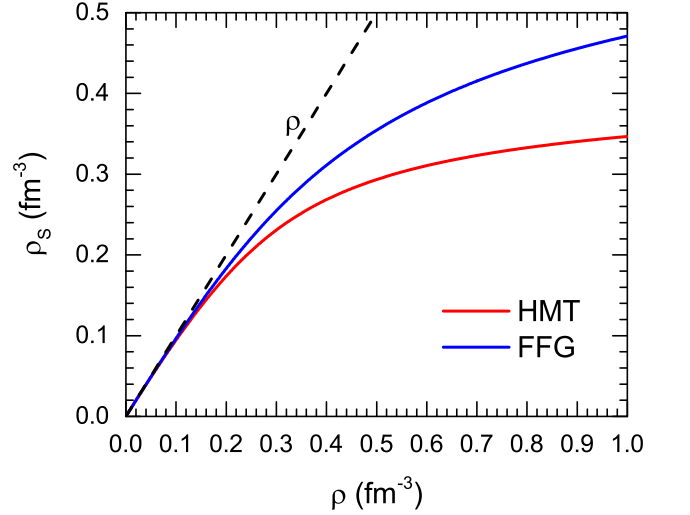}
  \caption{(Color Online) Scalar density of SNM as a function of baryon density for both the FFG and HMT models.}
  \label{fig-RHOSRMFHMT}
\end{figure}
As discussed earlier, the kinetic symmetry energy depends on the
nucleon Dirac effective mass which is determined by the scalar
baryon density $\rho_{\rm{S}}$. It is thus interesting to examine
explicitly how the SRC-modified nucleon momentum distribution
affects the scalar density and the Dirac effective mass. As shown in
Appendix B,  see (\ref{SNMRHOS}), the scalar density $\rho_{\rm{S}}$
can be written as
\begin{align}
\rho_{\rm{S}}=&\frac{\Delta_0M_0^{\ast,3}}{\pi^2}
\left(\theta\sqrt{1+\theta^2}-\rm{arcsinh}\,\theta\right)\notag\\
&+\frac{2C_0k_{\rm{F}}^4}{\pi^2M_0^{\ast}}\left(\sqrt{1+\frac{1}{\theta^2}}-\sqrt{1+\frac{1}{\phi_0^2\theta^2}}\right)
\end{align} with $\theta=k_{\rm{F}}/M_0^{\ast}$.  At low densities, the $\theta$
is small, thus
\begin{align}
\theta\sqrt{1+\theta^2}-\rm{arcsinh}\theta\approx\frac{2}{3}\theta^3-\frac{1}{5}\theta^5,\\
\theta^4\left(\sqrt{1+\frac{1}{\theta^2}}-\sqrt{1+\frac{1}{\phi_0^2\theta^2}}\right)&\notag\\\approx\left(1-\frac{1}{\phi_0}\right)\theta^3
+\frac{1}{2}(1-\phi_0)\theta^5.
\end{align}
Keeping only the first term, one has,
\begin{align}
\rho_{\rm{S}}\longrightarrow&\frac{\Delta_0M_0^{\ast,3}}{\pi^2}\frac{2}{3}\theta^3+\frac{2C_0M_0^{\ast,3}}{\pi^2}\left(1-\frac{1}{\phi_0}\right)\theta^3\notag\\
=&\frac{2M_0^{\ast,3}\theta^3}{3\pi^2}\left[\Delta_0+3C_0\left(1-\frac{1}{\phi_0}\right)\right]
=\rho,\end{align} and the next order correction to $\rho_{\rm{S}}$
is
\begin{equation}
\frac{M_0^{\ast,3}\theta^5}{\pi^2}\left[-\frac{1}{5}+C_0\left(\frac{8}{5}-\frac{3}{5\phi_0}-\phi_0\right)\right]
\end{equation}
which is negative, leading to $\rho_{\rm{S}}<\rho$. For the FFG
model ($\phi_0=1,\phi_1=0$), the value in the bracket of the above
expression is $-1/5$ while the term
$C_0\left({8}/{5}-{3}/{5\phi_0}-\phi_0\right)$ is always negative.
At the high density limit $\rho\to\infty$, the
$\sigma$ field will saturate at the value of
$\overline{\sigma}^{\infty}\equiv\overline{\sigma}(\rho=\infty)=M/g_{\sigma}$
(for the Dirac effective mass
$M_0^{\ast}=M-g_{\sigma}\overline{\sigma}$ approaches zero in
this limit). Correspondingly, we obtain
$\rho_{\rm{S}}^{\infty}\equiv\rho_{\rm{S}}(\rho=\infty)=M^3[(m_{\sigma}/g_{\sigma}M)^2+b_{\sigma}+c_{\sigma}]$
according to Eq. (\ref{eomf}). More quantitatively, we have
$\rho_{\rm{S}}^{\infty}(\rm{FFG})\approx3.29\,\rm{fm}^{-3}$ and
$\rho_{\rm{S}}^{\infty}(\rm{HMT})\approx2.99\,\rm{fm}^{-3}$.
Thus, the scalar density in the HMT model is always smaller than
that in the FFG model as shown in Fig. \ref{fig-RHOSRMFHMT}.

In Fig. \ref{fig-DiracRMFHMT}, the nucleon Dirac effective masses in SNM in the FFG and HMT models are
shown. The two models are found to give very similar results. This is easy to understand.
On one hand, three points of the effective mass are fixed, i.e., $M_0^{\ast}(0)/M=0$,
$M_0^{\ast}(\rho_0)/M=0.6$ and $M_0^{\ast}(\infty)/M=0$. On the
other hand, through Eq. (\ref{f0withrho}) we know that $\partial
\overline{\sigma}/\partial\rho>0$. Thus, the $M_0^{\ast}/M$ monotonically decreases in the
whole density range and is concave at large densities. Not surprisingly,  meeting all of these common constraints the $M_0^{\ast}(\rho)/M$
in the two models behaves very similarly.
\begin{figure}[h!]
\centering
  \includegraphics[width=8.5cm]{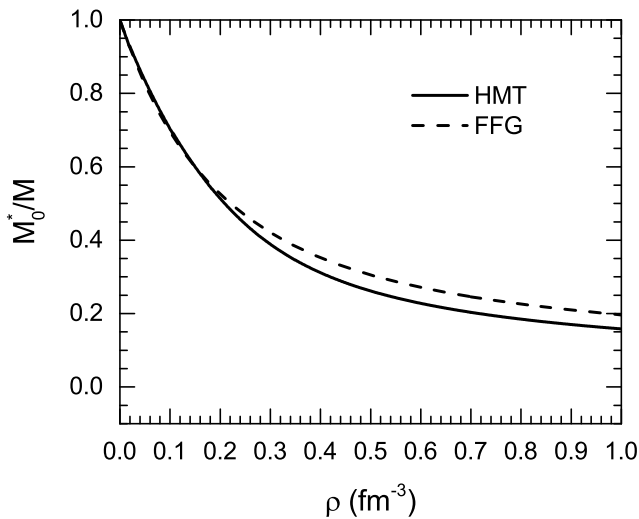}
  \caption{ Nucleon Dirac effective mass in SNM as a function of density for both the FFG and HMT models.}
  \label{fig-DiracRMFHMT}
\end{figure}

\section{The Total Symmetry Energy in the RMF Model with High Momentum Nucleons}\label{SecVI}
\begin{figure}[h!]
\centering
  \includegraphics[width=8.5cm]{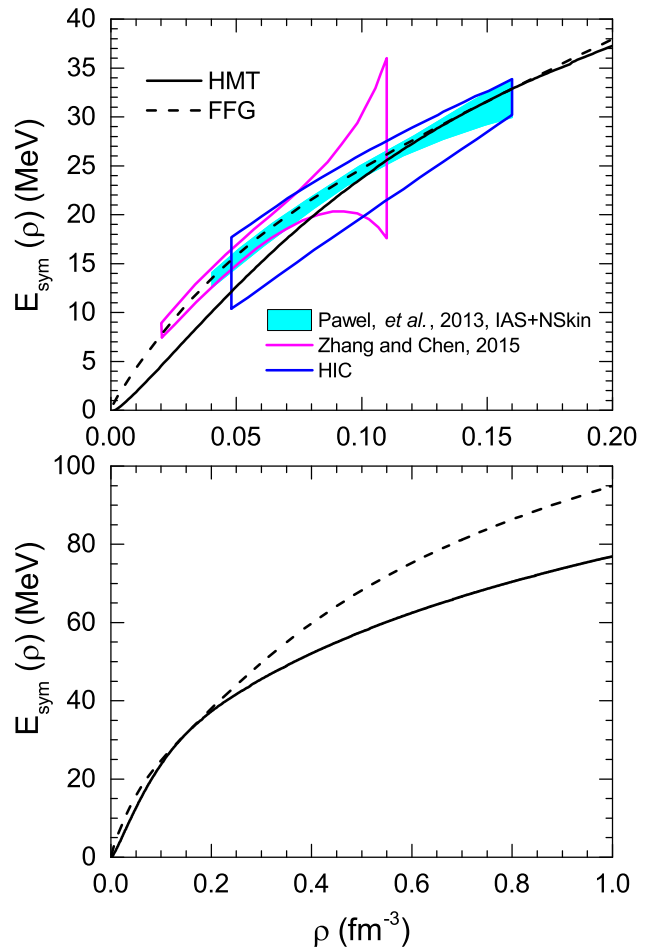}
  \caption{(Color Online) The total symmetry energy as a function of density in the HMT and FFG models in comparison with constraints
  from several recent studies\,\cite{Dan14,Tsa12,Zha15}.}
  \label{fig-EsymRMFHMT}
\end{figure}

We now turn to the total symmetry energy $E_{\rm{sym}}(\rho)$. Shown
in Fig. \ref{fig-EsymRMFHMT} are the HMT and FFG model predictions
in comparison with results from several recent studies by
others\,\cite{Dan14,Tsa12,Zha15}. It is seen from the upper panel
that the HMT softens the $E_{\rm{sym}}(\rho)$ at sub-saturation
densities. For instance, at densities around 0.04 fm$^{-3}$, the
effect is about 30\% which is larger than the width of the existing
constraint\,\cite{Zha15}.  Interestingly, at supra-saturation
densities as shown in the lower panel, the symmetry energy is also
significantly softened by the HMT. For instance, the effect is about
25\% at densities around 0.5 fm$^{-3}$. Thus, the HMT in the nucleon
momentum distribution provides a possible mechanism to soften the
symmetry energy at both low and high densities. Actually in the
original nonlinear RMF model, the high density $E_{\rm{sym}}(\rho)$
can not be made arbitrary small because of the structure of the
model itself. In the presence of HMT, however, mainly owing to the
negative kinetic symmetry energy it is possible that the total
$E_{\rm{sym}}(\rho)$ becomes very soft and even decreases at high
densities as indicated by some data analyses\,\cite{Xia09}.

Since the HMT and FFG models are designed to have the same values of
symmetry energy $E_{\rm{sym}}(\rho_0)$ and its slope $L$, it is
useful to use the curvature of the symmetry energy
\begin{align}
K_{\rm{sym}}\equiv\left[9\rho^2\frac{\partial^2E_{\rm{sym}}(\rho)}{\partial\rho^2}\right]_{\rho_0}
=\left[3\rho\frac{\partial L(\rho)}{\partial\rho}-3L(\rho)\right]_{\rho_0}
\end{align}
to measure the HMT effects on the total symmetry energy.
The $K_{\rm{sym}}$ is relevant for studying the isospin dependence of the incompressibility of ANM through the relationship
\begin{equation}
K(\delta)\approx K_0+K_{\tau}\delta^2+\mathcal{O}(\delta^4)
\end{equation}
where the $K_{\tau}$ is given by\,\cite{Che09}
\begin{equation}
K_{\tau}=K_{\rm{sym}}-6L-\frac{Q_0L}{K_0}.
\end{equation}
More quantitatively, we obtained the values of $
K_{\rm{sym}}^{\rm{FFG}}\approx-37\,\rm{MeV}$ and $
K_{\rm{sym}}^{\rm{HMT}}\approx-274\,\rm{MeV}$. The corresponding
isospin-coefficients of the incompressibility are
$K_{\tau}^{\rm{FFG}}\approx-174\,\rm{MeV}$ and
$K_{\tau}^{\rm{HMT}}\approx-470\,\rm{MeV}$. The latter is in very
good agreement with the best estimate of $K_{\tau}=-550\pm 100$ MeV
from analyzing many different kinds of experimental data currently
available\,\cite{Col14}. Overall, the HMT is to make the symmetry
energy significantly more concave around the saturation density,
leading to a stronger isospin dependence in the incompressibility of
ANM compared to calculations using the FFG model.

\section{Some Effects of High Momentum nucleons on properties of Neutron Stars}\label{SecVII}
The SRC-modified single-nucleon momentum distribution is expected to
affect some properties of neutron stars, see, e.g., ref.
\cite{Fra08}. First of all, the softening of the symmetry energy is
generally expected to reduce the proton fraction
$x_{\rm{p}}=\rho_{\rm{p}}/\rho$ in neutron stars within the
parabolic approximation of the EOS of ANM. For example, in the npe
matter at $\beta$ equilibrium, according to the chemical equilibrium
and charge neutrality conditions for reactions of
$\textrm{n}\to\textrm{p}+\textrm{e}+\overline{\nu}_{\textrm{e}}$ and
$ \textrm{p}+\textrm{e}\to\textrm{n}+\nu_{\textrm{e}}$, we have $
\mu_{\textrm{e}}=\mu_{\textrm{n}}-\mu_{\textrm{p}} $ where
$\mu_{\textrm{e}}=[m_{\textrm{e}}^2+(k_{\textrm{F}}^{\textrm{e}})^2]^{1/2}=[m_{\textrm{e}}^2+(3\pi^2\rho
x_{\textrm{e}})^{2/3}]^{1/2}\simeq (3\pi^2\rho
x_{\textrm{e}})^{1/3}$ with $x_{\textrm{e}}\equiv
\rho_{\textrm{e}}/\rho$ the electron fraction, i.e.,
$\mu_{\rm{e}}=\mu_{\rm{n}}-\mu_{\rm{p}}\approx
4E_{\rm{sym}}(\rho)\delta+\mathcal{O}(\delta^3)$. Thus, the
$x_{\rm{p}}=x_{\rm{e}}$ will be reduced if the symmetry energy
$E_{\rm{sym}}(\rho)$ decreases. However, we caution that although
high order terms in the EOS of ANM are relatively small, they still
have non-negligible effects on the $x_{\rm{p}}$ in the RMF
models\,\cite{Cai12}. Expectations based on the parabolic
approximation for the EOS of ANM may be altered. Moreover, it was
suggested recently in ref. \cite{Don15} that the neutrino emissivity
of the direct URCA process will be reduced by a factor
$\eta=Z_{\rm{F}}^{\rm{p}}Z_{\rm{F}}^{\rm{n}}$ compared to the FFG
model with $Z_{\rm{F}}^J$ the discontinuity of the single-nucleon
momentum distribution at the Fermi momentum. In the HMT model, the
depletion of the Fermi sphere together with the sizable value of
$C_J$ makes the factor $\eta$ much smaller than unity. However,
effects of nucleons in the HMT not considered in ref. \cite{Don15}
may enhance the emissivity of neutrinos \cite{Fra08,Alex15}. Thus,
to our best knowledge, the net effects of the entire single-nucleon
momentum distribution modified by the SRC on both the critical
density for the direct URCA process to occur and the cooling rate of
protoneutron stars are still unclear. Nevertheless, it is
interesting to note that efforts to clarify the issue are currently
underway \cite{Will}.

\begin{figure}[h!]
\centering
  \includegraphics[width=8.cm]{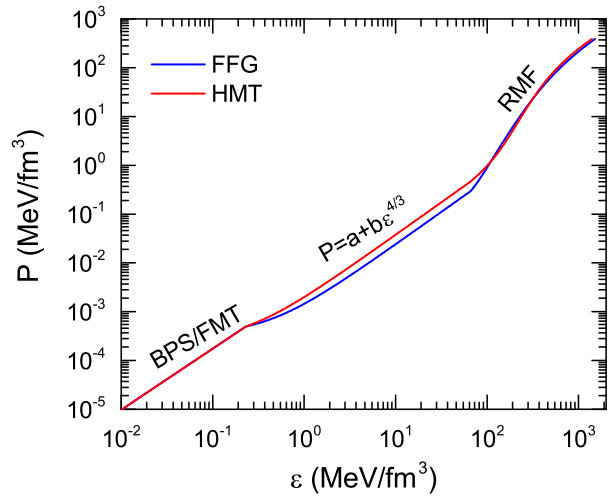}
  \caption{(Color Online) EOS of neutron star matter. Detailed descriptions of the compositions of different layers of the neutron stars are explained in the text.}
  \label{fig-betaEoS}
\end{figure}

Next, we investigate effects of the HMT on the mass-radius relation of
neutron stars. In constructing the EOS of various layers in neutron
stars for solving the Tolman-Oppenheimer-Volkoff (TOV) equation, we
follow a standard scheme. Neutron stars are composed of the $npe$ matter at low densities as described above. For the core
we use the EOS of $\beta$-stable and charge neutral npe$\mu$ matter
obtained from the nonlinear RMF model described earlier. When the
chemical potential of electron is larger than the static mass of a
muon, reactions $
\textrm{e}\to\mu+\nu_{\textrm{e}}+\overline{\nu}_{\mu}$, $
\textrm{p}+\mu\to\textrm{n}+\nu_{\mu}$ and
$\textrm{n}\to\textrm{p}+\mu+\overline{\nu}_{\mu}$ will also take
place. The latter requires
\begin{equation}\label{ceq2} \mu_{\textrm{n}}-\mu_{\textrm{p}}=\mu_{\mu}=\sqrt{m_{\mu}^2+(3\pi^2\rho
x_{\mu})^{2/3}}
\end{equation} besides
$\mu_{\textrm{n}}-\mu_{\textrm{p}}=\mu_{\textrm{e}}$, where
$m_{\mu}=105.7\,\textrm{MeV}$ is the mass of a muon and
$x_{\mu}\equiv \rho_{\mu}/\rho$ is the muon fraction. The inner
crust with densities ranging between $\rho_{\text{out}}=2.46\times
10^{-4}$ fm$^{-3}$ corresponding to the neutron dripline and the
core-crust transition density $\rho _{\text{t}}$ is the region where
some complex and exotic structures\,---\,collectively referred to
as the ``nuclear pasta" may exist.  Because of our poor knowledge
about this region we adopt the polytropic EOSs parameterized in
terms of the pressure $P$ as a function of total energy density
$\varepsilon$ according to
$P=a+b\varepsilon^{4/3}$\,\cite{XuJ09,Hor03}. The constants $a$ and
$b$ are determined by the pressure and energy density at $\rho
_{\text{t}}$ and $\rho _{\text{out}}$\,\cite{XuJ09}. For the outer
crust\,\cite{BPS71}, we use the BPS EOS for the region with
$6.93\times 10^{-13}$\,fm$^{-3}<\rho <\rho _{\text{out}}$ and the
FMT EOS for $4.73\times 10^{-15}$\,fm$^{-3}<\rho <$$6.93\times
10^{-13}$\,fm$^{-3}$, respectively.
\begin{figure}[h!]
\centering
  \includegraphics[width=8.cm]{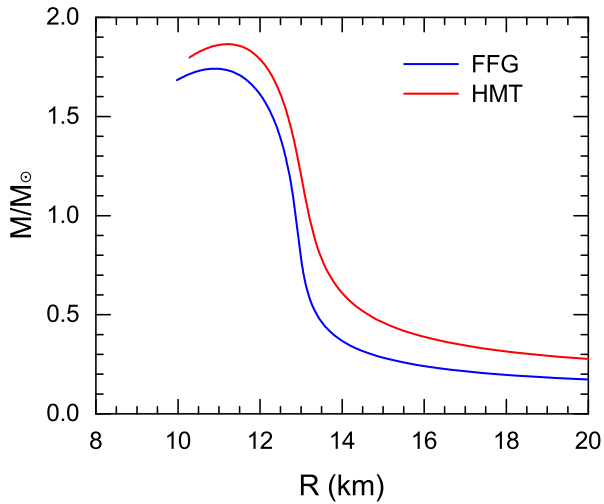}
  \caption{(Color Online) Mass-radius of a neutron star obtained by integrating the TOV equation under the EOS of neutron star matter in the FFG and HMT models, respectively.}
  \label{fig-MR}
\end{figure}

Shown in Fig. \ref{fig-betaEoS} are the EOSs of neutron star matter
obtained within the FFG and HMT models. The similarity of the two model EOSs indicates
that the corresponding mass-radius relations will not be different
dramatically. In Fig. \ref{fig-MR}, the corresponding mass-radius relations of
neutron stars from the two model are compared. As discussed earlier in Section \ref{SecIV}, the
skewness of SNM mainly characterizes the high density behavior
of the EOS of SNM. The SRC induced HMT is to increase the skewness of SNM and thus hardens the EOS of neutron star matter.
On the other hand, the symmetry energy effect on the mass-radius relation of neutron stars is
relatively smaller in the RMF models\,\cite{Ser79}. Therefore, the enhanced skewness $Q_0$
due to the HMT enlarges the maximum mass of neutron stars as shown in the Fig. \ref{fig-MR}. Quantitatively,
the maximum mass of neutron stars in the HMT and FFG models are $M^{\rm{max}}\approx1.87\,M_{\odot}$ and
$M^{\max}\approx1.74\,M_{\odot}$, with the corresponding radii being
about $11.21\,\rm{km}$ and $10.89\,\rm{km}$, respectively. The
relative effect on the maximum mass is about $8\%$. While the maximum mass is still below the observational data,
the EOS with HMT helps improve the situation.

\section{Summary and Remaks}\label{SecVIII}

In summary, within the nonlinear RMF model incorporating the
SRC-modified single-nucleon momentum distribution constrained by
findings of recent electron-nucleus scattering experiments, we have
studied the EOS of asymmetric nucleonic matter. In particular, the
kinetic symmetry energy in the presence of SRC-induced high momentum
nucleons is found to be
$E_{\rm{sym}}^{\rm{kin}}(\rho_0)=-16.94\pm13.66\,\rm{MeV}$
consistent with earlier findings in non-relativistic models. Similar
to the findings about the nucleon spectroscopic factors, the average
nucleon kinetic energy extracted from electron-nucleus scattering
experiments can not be reproduced by traditional RMF models lacking
correlations. Including the SRC-induced high momentum nucleons in
the RMF model, the data can be well reproduced. Comparing the RMF
calculations with and without the SRC-induced high momentum nucleons
using two sets of model parameters both reproducing identically all
empirical properties of SNM and the symmetry energy of ANM at
$\rho_0$, the SRC-modified single-nucleon momentum distribution is
found to make the EOS of SNM much harder at supra-saturation densities and the
$E_{\rm{sym}}(\rho)$ more concave around $\rho_0$, leading to a larger maximum mass of neutron stars
and an isospin-dependent incompressibility of ANM  in better agreement with
existing observational/experimental data.

After introducing the SRC-induced high momentum nucleons, some
isovecor parameters of the RMF model had to be readjusted to
reproduce the same empirical properties of ANM and known
experimental constraints. Ramifications of these changes and the
resulting symmetry energy on experimental observables, such as
neutron skins of heavy nuclei will be studied in the near future. The SRC and some of its effects in
nuclear structures and reactions are well established both
theoretically and experimentally. While the RMF model has been very
successful in helping us understand many fundamental physics and
explaining various experimental phenomena, as a mean-field model by
design it lacks correlations that are important and necessary in
understanding some other experimental phenomena. Going beyond the
mean-field level, we take one step forward by replacing the step
function with an experimentally constrained momentum distribution
incorporating the SRC-induced HMT in reformulating some aspects of
the nonlinear RMF model. Compared with fully microscopic many-body
theories where effects of the SRC are considered self consistently,
our hybrid approach is relatively simple but transparent and all relevant
physical quantities are given analytically. While much more work
remains to be done, the analyses and numerical results presented
here are instructive for better understanding effects of the SRC on
the EOS of neutron-rich nucleonic matter which is relevant for both
nuclear physics and astrophysics.

\section*{Acknowledgement}
We would like to thank William G. Newton and Isaac Vida\~na for helpful discussions.
This work was supported in part by the U.S. National Science Foundation under Grant No. PHY-1068022, the U.S. Department of Energy's Office of Science under Award Number DE-SC0013702
and the National Natural Science Foundation of China under grant no. 11320101004.

\appendix
\section{The Derivation of $E_{\textmd{sym}}^{\textmd{kin}}(\rho)$}\label{App1}

We start from the kinetic energy density of (\ref{EnDenKin})
\begin{align}
\varepsilon_J^{\rm{kin}}
=&\frac{\Delta_J}{\pi^2}\int_0^{k_J}k^2\sqrt{k^2+m_J^2}\d
k\notag\\
&+\frac{C_Jk_J^4}{\pi^2}\int_{k_J}^{\phi_Jk_J}\frac{1}{k^2}\sqrt{k^2+m_J^2}\d
k.
\end{align}
Using the abbreviations of $ m_J\equiv
M^{\ast}_J,k_J=k_{\rm{F}}^J,\xi=k/m_J,p_J=\phi_J k_J$, the kinetic
energy density can be expressed as
\begin{align}
\varepsilon_J^{\rm{kin}}=&\frac{\Delta_Jm_J^4}{\pi^2}\int_0^{k_J/m_J}\xi^2\sqrt{1+\xi^2}\d\xi\notag\\
&+\frac{C_Jk_J^4}{\pi^2}\int_{k_J/m_J}^{\phi_Jk_J/m_J}\frac{\sqrt{1+\xi^2}}{\xi^2}\d\xi.
\end{align}
In order to use the following elementary formula,
\begin{equation}
\frac{\d}{\d x}\int^{w(x)}\d yz(y)=z(w(x))\frac{\d w(x)}{\d x},
\end{equation}
we rewrite the kinetic energy density as
\begin{widetext}
\begin{equation}
\varepsilon_J^{\rm{kin}}=\frac{\Delta_Jm_J^4}{\pi^2}\int_0^{k_J/m_J}\xi^2\sqrt{1+\xi^2}\d\xi
+\frac{C_Jk_J^4}{\pi^2}\left[\int_{0}^{p_J/m_J}\frac{\sqrt{1+\xi^2}}{\xi^2}\d\xi-
\int_{0}^{k_J/m_J}\frac{\sqrt{1+\xi^2}}{\xi^2}\d\xi\right],
\end{equation}
then its derivative with respect to the isospin asymmetry $\delta$ is
\begin{align}
\varepsilon_J^{\rm{kin}'}=&\frac{4m_J'\Delta_Jm_J^3+\Delta_J'm_J^4}{\pi^2}\cdot\int_0^{k_J/m_J}\xi^2\sqrt{1+\xi^2}\d\xi
+\frac{k_J'\Delta_Jk_J^2E_J^{\ast}}{\pi^2}-\frac{m_J'\Delta_Jk_J^3E_J^{\ast}}{\pi^2m_J}\notag\\
&+\frac{4k_J'C_Jk_J^3+C_J'k_J^4}{\pi^2}\cdot\int_{k_J/m_J}^{p_J/m_J}\frac{\sqrt{1+\xi^2}}{\xi^2}\d\xi
+\frac{C_Jk_J^4}{\pi^2}\cdot\left(\frac{p_J'F_J^{\ast}}{p_J^2}-\frac{m_J'F_J^{\ast}}{p_Jm_J}-
\frac{k_J'E_J^{\ast}}{k_J^2}+\frac{m_J'E_J^{\ast}}{k_Jm_J}\right)
\end{align}
where
$F_J^{\ast}=(p_J^2+m_J^2)^{1/2}=(\phi_J^2k_J^2+m_J^2)^{1/2}$.

The second order derivative of $\varepsilon_J^{\rm{kin}}$ with respect to $\delta$ can be obtained
in a similar way, i.e.,
\begin{align}
\varepsilon_J^{\rm{kin}''}=&\frac{1}{\pi^2}\left(8\Delta_J'm_J'm_J^3+12m_J'^2\Delta_Jm_J^2+4m_J''\Delta_Jm_J^3
+\Delta_J''m_J^4\right)\int_0^{k_J/m_J}\xi^2\sqrt{1+\xi^2}\d\xi\notag\\
&+\frac{1}{\pi^2m_J}\Big(m_J'k_J'\Delta_Jk_J^2E_J^{\ast}
+\Delta_J'k_J'k_J^2m_JE_J^{\ast}-2\Delta_J'm_J'k_J^3E_J^{\ast}
-m_J'E_J^{\ast'}\Delta_Jk_J^3-m_J''E_J^{\ast}\Delta_Jk_J^3\Big)\notag\\
&+\frac{1}{\pi^2}\left(\Delta_J'k_J'k_J^2E_J^{\ast}+2k_J'^2\Delta_Jk_JE_J^{\ast}
+k_J'E_J^{\ast'}\Delta_Jk_J^2+k_J''\Delta_Jk_J^2E_J^{\ast}\right)-\frac{3m_J'^2\Delta_Jk_J^3E_J^{\ast}}{\pi^2m_J^2}\notag\\
&+\frac{1}{\pi^2}\left(8C_J'k_J'k_J^3+12k_J'^2C_Jk_J^2+4k_J''C_Jk_J^3+C_J''k_J^4\right)
\int_{k_J/m_J}^{p_J/m_J}\frac{\sqrt{1+\xi^2}}{\xi^2}\d\xi\notag\\
&+\frac{8k_J'C_Jk_J^3+2C_J'k_J^4}{\pi^2}\cdot\left(\frac{p_J'F_J^{\ast}}{p_J^2}-\frac{m_J'F_J^{\ast}}{p_Jm_J}-
\frac{k_J'E_J^{\ast}}{k_J^2}+\frac{m_J'E_J^{\ast}}{k_Jm_J}\right)\notag\\
&+\frac{C_Jk_J^4}{\pi^2}\cdot\Bigg(\frac{F_J^{\ast'}p_J'}{p_J^2}+\frac{F_J^{\ast}p_J''}{p_J^2}-\frac{2F_J^{\ast}p_J'^2}{p_J^3}
-\frac{F_J^{\ast'}m_J'}{p_Jm_J}-\frac{F_J^{\ast}m_J''}{p_Jm_J}
+\frac{F_J^{\ast}m_J'p_J'}{p_J^2m_J}+\frac{F_J^{\ast}m_J'^2}{p_Jm_J^2}\Bigg)\notag\\
&-\frac{C_Jk_J^4}{\pi^2}\cdot\Bigg(\frac{E_J^{\ast'}k_J'}{k_J^2}+\frac{E_J^{\ast}k_J''}{k_J^2}-\frac{2E_J^{\ast}k_J'^2}{k_J^3}
-\frac{E_J^{\ast'}m_J'}{k_Jm_J}-\frac{E_J^{\ast}m_J''}{k_Jm_J}
+\frac{E_J^{\ast}m_J'k_J'}{k_J^2m_J}+\frac{E_J^{\ast}m_J'^2}{k_Jm_J^2}\Bigg).
\end{align}

We introduce the abbreviations $ f\equiv\overline{\sigma},
w\equiv\overline{\omega}_0$, and a subscript ``0" denotes the
symmetric case, for example, $f_0\equiv
f(\delta=0),\left.f_0'={\partial
f}/{\partial\delta}\right|_{\delta=0}$, etc.  The scalar density is
a function of $m_J$, i.e., $\rho_{\rm{S},J}=\rho_{\rm{S},J}(m_J)$,
then $\rho_{\rm{S},J}'=\lambda_Jm_J'$ with $\lambda_J$ a certain
factor. From the field equation of $f$, it is easy to find that
$f_0'=0$. This means that we can omit the terms proportional to
$m_J'$ in the expression of $\varepsilon_J^{\rm{kin}''}$ (since the
symmetry energy is obtained by taking $\delta=0$ in
$\varepsilon_J^{\rm{kin}''}$), thus we have (omitting the
terms proportional to $m_J'$)
\begin{align}
\varepsilon_J^{\rm{kin}''}=&\frac{1}{\pi^2}\left(4m_J''\Delta_Jm_J^3
+\Delta_J''m_J^4\right)\cdot\int_0^{k_J/m_J}\xi^2\sqrt{1+\xi^2}\d\xi
+\frac{1}{\pi^2m_J}\left(\Delta_J'k_J'k_J^2m_JE_J^{\ast}-m_J''E_J^{\ast}\Delta_Jk_J^3\right)\notag\\
&+\frac{1}{\pi^2}\left(\Delta_J'k_J'k_J^2E_J^{\ast}+2k_J'^2\Delta_Jk_JE_J^{\ast}
+k_J'E_J^{\ast'}\Delta_Jk_J^2+k_J''\Delta_Jk_J^2E_J^{\ast}\right)+\frac{8k_J'C_Jk_J^3+2C_J'k_J^4}{\pi^2}\cdot\left(\frac{p_J'F_J^{\ast}}{p_J^2}-
\frac{k_J'E_J^{\ast}}{k_J^2}\right)\notag\\
&+\frac{1}{\pi^2}\left(8C_J'k_J'k_J^3+12k_J'^2C_Jk_J^2+4k_J''C_Jk_J^3+C_J''k_J^4\right)
\int_{k_J/m_J}^{p_J/m_J}\frac{\sqrt{1+\xi^2}}{\xi^2}\d\xi\notag\\
&+\frac{C_Jk_J^4}{\pi^2}\cdot\left(\frac{F_J^{\ast'}p_J'}{p_J^2}+\frac{F_J^{\ast}p_J''}{p_J^2}-\frac{2F_J^{\ast}p_J'^2}{p_J^3}
-\frac{F_J^{\ast}m_J''}{p_Jm_J}\right)
-\frac{C_Jk_J^4}{\pi^2}\cdot\left(\frac{E_J^{\ast'}k_J'}{k_J^2}+\frac{E_J^{\ast}k_J''}{k_J^2}-\frac{2E_J^{\ast}k_J'^2}{k_J^3}
-\frac{E_J^{\ast}m_J''}{k_Jm_J}\right).
\end{align}

We then deal with the terms proportional to the second derivative of
$m_J''$, i.e.,
\begin{align}
\Pi=m_J''\Bigg[&\frac{4}{m_J}\frac{\Delta_J
m_J^4}{\pi^2}\int_0^{k_J/m_J}\xi^2\sqrt{1+\xi^2}\d\xi
-\frac{\Delta_Jk_J^3E_J^{\ast}}{\pi^2m_J}
-\frac{C_Jk_J^4}{\pi^2}\left(\frac{F_J^{\ast}}{p_Jm_J}-\frac{E_J^{\ast}}{k_Jm_J}\right)\Bigg]
.
\end{align}
Moreover,
\begin{align}
V=&\int_0^{k_J/m_J}\xi^2\sqrt{1+\xi^2}\d\xi=\frac{1}{4}x(x^2+1)^{3/2}-\frac{1}{8}x(x^2+1)^{1/2}-\frac{1}{8}\rm{arcsinh}\,x,\\
S=&\int_0^{k_J/m_J}\frac{\xi^2\d\xi}{\sqrt{1+\xi^2}}=\frac{1}{2}x(x^2+1)^{1/2}-\frac{1}{2}\rm{arcsinh}\,x,
\end{align}
with $x=k_J/m_J$. For $k<k_J$, we then have
\begin{align} \varepsilon_J^{\rm{kin,I}}&=\frac{\Delta_J
m_J^4}{\pi^2}\int_0^{k_J/m_J}\xi^2\sqrt{1+\xi^2}\d\xi=\frac{\Delta_J
m_J^4V}{\pi^2},~~
\rho_{\rm{S},J}^{\rm{I}}=\frac{\Delta_Jm_J^3}{\pi^2}\int_0^{k_J/m_J}\frac{\xi^2\d\xi}{\sqrt{1+\xi^2}}=\frac{\Delta_Jm_J^3S}{\pi^2},
\end{align}
so
\begin{equation}
\varepsilon_J^{\rm{kin,I}}=\frac{\Delta_Jk_J^3E_J^{\ast}}{4\pi^2}+\frac{m_J}{4}\rho_{\rm{S},J}^{\rm{I}}.\end{equation}
On the other hand, the high momentum part of the scalar density
should be written as
\begin{align}
\rho_{\rm{S},J}^{\rm{II}}=&\frac{C_Jk_J^4}{\pi^2m_J}\int_{k_J/m_J}^{p_J/m_J}\frac{\d\xi}{\xi^2\sqrt{1+\xi^2}}
=-\frac{C_Jk_J^4}{\pi^2m_J}\left(\frac{F_J^{\ast}}{p_J}-\frac{E_J^{\ast}}{k_J}\right)
.\end{align} Then $\Pi$ should be rewritten as
\begin{equation}
\Pi=m_J''\left[\frac{4}{m_J}\left(\frac{\Delta_Jk_J^3E_J^{\ast}}{4\pi^2}+\frac{m_J}{4}\rho_{\rm{S},J}^{\rm{I}}\right)-\frac{\Delta_Jk_J^3E_J^{\ast}}{\pi^2m_J}
+\rho_{\rm{S},J}^{\rm{II}}\right]=\rho_{\rm{S},J}m_J''.\end{equation}
After combining this term with the corresponding potential terms in
the energy density, it is found that there exist no terms
proportional to $f_0''$ according to the equation of motion of $f$,
thus it is safe to write $\varepsilon_J^{\rm{kin}''}$ as (omitting
the terms proportional to $m_J''$)
\begin{align}
\varepsilon_J^{\rm{kin}''}=&\frac{\Delta_J''m_J^4}{\pi^2}\int_0^{k_J/m_J}\xi^2\sqrt{1+\xi^2}\d\xi
+\frac{1}{\pi^2}\left(2\Delta_J'k_J'k_J^2E_J^{\ast}+2k_J'^2\Delta_Jk_JE_J^{\ast}
+k_J'E_J^{\ast'}\Delta_Jk_J^2+k_J''\Delta_Jk_J^2E_J^{\ast}\right)\notag\\
&+\frac{1}{\pi^2}\left(8C_J'k_J'k_J^3+12k_J'^2C_Jk_J^2+4k_J''C_Jk_J^3\right)\cdot\int_{k_J/m_J}^{p_J/m_J}\frac{\sqrt{1+\xi^2}}{\xi^2}\d\xi\notag\\
&+\frac{8k_J'C_Jk_J^3+2C_J'k_J^4}{\pi^2}\cdot\left(\frac{p_J'F_J^{\ast}}{p_J^2}-
\frac{k_J'E_J^{\ast}}{k_J^2}\right)\notag\\
&+\frac{C_Jk_J^4}{\pi^2}\cdot\left(\frac{F_J^{\ast'}p_J'}{p_J^2}+\frac{F_J^{\ast}p_J''}{p_J^2}-\frac{2F_J^{\ast}p_J'^2}{p_J^3}
\right)
-\frac{C_Jk_J^4}{\pi^2}\cdot\left(\frac{E_J^{\ast'}k_J'}{k_J^2}+\frac{E_J^{\ast}k_J''}{k_J^2}-\frac{2E_J^{\ast}k_J'^2}{k_J^3}
\right),\label{F1}
\end{align}
where the terms proportional to $C_J''$ are also omitted for  $C_J$
is linear in $\delta$, i.e., $C_J''=0$.
\end{widetext}

For \begin{equation}Y_J=Y_0(1+Y_1\tau_3^J\delta)\end{equation} with
$Y=C,\phi$, we have \begin{equation}
Y_J'=Y_0Y_1\tau_3^J.\end{equation} Thus
\begin{align}
\Delta_J'=&-3C_J'\left(1-\frac{1}{\phi_J}\right)-\frac{3C_J\phi_J'}{\phi_J^2}\notag\\
&\longrightarrow-3C_0\left[C_1\left(1-\frac{1}{\phi_0}\right)+\frac{\phi_1}{\phi_0}\right]\tau_3^J,\\
\Delta_J''=&-\frac{6C_J'\phi_J'}{\phi_J^2}+\frac{6C_J\phi_J'^2}{\phi_J^3}\notag\\
&\longrightarrow-\frac{6C_0\phi_1}{\phi_0}\left(C_1-\phi_1\right),
\end{align}
and
\begin{align}
k_J'=&\frac{1}{3}k_{\rm{F}}\tau_3^J(\alpha\delta+\cdots)\longrightarrow\frac{1}{3}k_{\rm{F}}\tau_3^J,\\
k_J''=&-\frac{2}{9}k_{\rm{F}}(\beta\delta+\cdots)\longrightarrow-\frac{2}{9}k_{\rm{F}},\\
p_J'=&\phi_J'k_J+\phi_Jk_J'\longrightarrow\left(\phi_1+\frac{1}{3}\right)\phi_0k_{\rm{F}}\tau_3^J,\\
p_J''=&2\phi_J'k_J'+\phi_Jk_J''\longrightarrow\frac{2}{3}\phi_0k_{\rm{F}}\left(\phi_1-\frac{1}{3}\right),\\
E_J^{\ast'}=&\frac{m_Jm_J'+k_Jk_J'}{E_J^{\ast}}\longrightarrow\frac{k_{\rm{F}}^2}{3E_{\rm{F}}^{\ast}}\tau_3^J,\\
F_J^{\ast'}=&\frac{m_Jm_J'+p_Jp_J'}{F_J^{\ast}}\longrightarrow\frac{\phi_0^2k_{\rm{F}}^2(1+3\phi_1)}{3F_{\rm{F}}^{\ast}}\tau_3^J,
\end{align}
where ``$\longrightarrow$" means the limit of $\delta=0$. It is
clear that in the FFG model, $\phi_0=1,\phi_0=1$, then
\begin{equation}\Delta_J\longrightarrow1,~~\Delta_J',\Delta_J''\longrightarrow0,\end{equation} and
\[
\varepsilon_J^{\rm{kin}''}=\frac{1}{\pi^2}\left(2k_J'^2k_JE_J^{\ast}+k_J'E_J^{\ast'}k_J^2+k_J''k_J^2E_J^{\ast}\right),\]
which is expected. Evaluating
$\sum_{J=\rm{n,p}}\varepsilon_{J}^{\rm{kin}''}|_{\delta=0}$
according to (\ref{F1}) and then dividing it by $2\rho$, we shall obtain
(\ref{EsymkinHMT}).

\section{The Derivation of $L(\rho)$}\label{App2}

In order to derive the expressions for $L$, we should first obtain
an expression for $\partial f_0/\partial\rho$, which is related to
the scalar density that can be decomposed into two terms,
i.e.,\begin{widetext}\begin{equation}{
\rho_{\rm{S}}=\frac{2\Delta_0}{\pi^2}\int_0^{k_{\rm{F}}}\frac{k^2\d
kM_0^{\ast}}{\sqrt{{k}^2+M_0^{\ast,2}}}+\frac{2C_0k_{\rm{F}}^4}{\pi^2}\int_{k_{\rm{F}}}^{p_{\rm{F}}}\frac{1}{k^4}\frac{k^2\d
kM_0^{\ast}}{\sqrt{{k}^2+M_0^{\ast,2}}},}\end{equation} where
$M_0^{\ast}=M-g_{\sigma}f_0$. Putting $k=\zeta M_0^{\ast}$, then
\begin{equation}
\rho_{\rm{S}}=\rho_{\rm{S}}^{\rm{I}}+\rho_{\rm{S}}^{\rm{II}}=
\frac{2\Delta_0M_0^{\ast,3}}{\pi^2}\int_0^{k_{\rm{F}}/M_0^{\ast}}\frac{\zeta^2\d\zeta}{\sqrt{1+\zeta^2}}
+\frac{2C_0k_{\rm{F}}^4}{\pi^2M_0^{\ast}}\int_{k_{\rm{F}}/M_0^{\ast}}^{p_{\rm{F}}/M_0^{\ast}}
\frac{\d\zeta}{\zeta^2\sqrt{1+\zeta^2}},\end{equation} so
\begin{align}
\frac{\partial\rho_{\rm{S}}^{\rm{I}}}{\partial\rho}&=\frac{6\Delta_0M_0^{\ast,2}}{\pi^2}
\frac{\partial
M_0^{\ast}}{\partial\rho}\int_0^{k_{\rm{F}}/M_0^{\ast}}\frac{\zeta^2\d\zeta}{\sqrt{1+\zeta^2}}
+\frac{2\Delta_0M_0^{\ast,3}}{\pi^2}\frac{(k_{\rm{F}}/M_0^{\ast})^2}{\sqrt{1+(k_{\rm{F}}/M_0^{\ast})^2}}
\frac{\partial}{\partial\rho}\frac{k_{\rm{F}}}{M_0^{\ast}}
\notag\\
&=\frac{3}{M_0^{\ast}}\frac{\partial M_0^{\ast}}{\partial\rho}
\frac{2\Delta_0M_0^{\ast,3}}{\pi^2}\int_0^{k_{\rm{F}}/M_0^{\ast}}\frac{\zeta^2\d\zeta}{\sqrt{1+\zeta^2}}
+\frac{2\Delta_0M_0^{\ast,2}k_{\rm{F}}^2}{\pi^2{E}_{\rm{F}}^{\ast}}
\frac{\partial}{\partial\rho}\frac{k_{\rm{F}}}{M_0^{\ast}}\notag\\
&=-\frac{3g_{\sigma}\rho_{\rm{S}}^{\rm{I}}}{M_0^{\ast}}\frac{\partial
f_0}{\partial\rho} +
\frac{2\Delta_0M_0^{\ast,2}k_{\rm{F}}^2}{\pi^2{E}_{\rm{F}}^{\ast}}\left(\frac{\pi^2}{2M_0^{\ast}k_{\rm{F}}^2}+\frac{g_{\sigma}k_{\rm{F}}}{M_0^{\ast,2}}
\frac{\partial f_0}{\partial\rho}\right)\notag\\
&=-\frac{3g_{\sigma}\rho_{\rm{S}}^{\rm{I}}}{M_0^{\ast}}\frac{\partial
f_0}{\partial\rho}+\frac{\Delta_0M_0^{\ast}}{{E}_{\rm{F}}^{\ast}}+\frac{3\Delta_0g_{\sigma}\rho}{{E}_{\rm{F}}^{\ast}}\frac{\partial
f_0}{\partial\rho}
=-3g_{\sigma}\left(\frac{\rho_{\rm{S}}^{\rm{I}}}{M_0^{\ast}}-\frac{\Delta_0\rho}{{E}_{\rm{F}}^{\ast}}\right)\frac{\partial
f_0}{\partial\rho}+\frac{\Delta_0M_0^{\ast}}{{E}_{\rm{F}}^{\ast}}.
\end{align}
The following two relations are useful in the derivations,
\begin{equation}
\frac{\partial
k_{\rm{F}}}{\partial\rho}=\frac{\pi^2}{2k_{\rm{F}}^2},~~\frac{\partial
{E}_{\rm{F}}^{\ast}}{\partial\rho}=\frac{\pi^2}{2k_{\rm{F}}{E}_{\rm{F}}^{\ast}}-\frac{g_{\sigma}M_0^{\ast}}{{E}_{\rm{F}}^{\ast}}\frac{\partial
f_0}{\partial\rho}. \end{equation} Similarly,
\begin{align}
\frac{\partial\rho_{\rm{S}}^{\rm{II}}}{\partial\rho}
=&\frac{2C_0}{\pi^2}\frac{\partial}{\partial\rho}\frac{k_{\rm{F}}^4}{M_0^{\ast}}
\int_{k_{\rm{F}}/M_0^{\ast}}^{p_{\rm{F}}/M_0^{\ast}}
\frac{\d\zeta}{\zeta^2\sqrt{1+\zeta^2}}+\frac{2C_0k_{\rm{F}}^4}{\pi^2M_0^{\ast}}\frac{\partial}{\partial\rho}
\int_{k_{\rm{F}}/M_0^{\ast}}^{p_{\rm{F}}/M_0^{\ast}}
\frac{\d\zeta}{\zeta^2\sqrt{1+\zeta^2}}\notag\\
=&\left(\frac{4}{k_{\rm{F}}}\frac{\partial
k_{\rm{F}}}{\partial\rho}-\frac{1}{M_0^{\ast}}\frac{\partial
M_0^{\ast}}{\partial\rho}\right)\frac{2C_0k_{\rm{F}}^4}{\pi^2M_0^{\ast}}\int_{k_{\rm{F}}/M_0^{\ast}}^{p_{\rm{F}}/M_0^{\ast}}
\frac{\d\zeta}{\zeta^2\sqrt{1+\zeta^2}}
+\frac{2C_0k_{\rm{F}}^4}{\pi^2M_0^{\ast}}\left[\frac{M_0^{\ast,3}}{p_{\rm{F}}^2F_{\rm{F}}^{\ast}}\frac{\partial}{\partial\rho}\left(
\frac{p_{\rm{F}}}{M_0^{\ast}}\right)
-\frac{M_0^{\ast,3}}{k_{\rm{F}}^2E_{\rm{F}}^{\ast}}\left(
\frac{k_{\rm{F}}}{M_0^{\ast}}\right)\right]\notag\\
=&\frac{4\rho_{\rm{S}}^{\rm{II}}}{3\rho}+C_0k_{\rm{F}}^2M_0^{\ast}\left(\frac{\phi_0}{p_{\rm{F}}^2F_{\rm{F}}^{\ast}}
-\frac{1}{k_{\rm{F}}^2E_{\rm{F}}}\right)
+g_{\sigma}\left[\frac{\rho_{\rm{S}}^{\rm{II}}}{M_0^{\ast}}
+\frac{2C_0k_{\rm{F}}^4}{\pi^2}\left(\frac{1}{p_{\rm{F}}F_{\rm{F}}^{\ast}}-\frac{1}{k_{\rm{F}}E_{\rm{F}}^{\ast}}\right)\right]\frac{\partial
f_0}{\partial\rho}.
\end{align}
Introducing,
\begin{equation}
\Phi=C_0k_{\rm{F}}^2M_0^{\ast}\left(\frac{\phi_0}{p_{\rm{F}}^2F_{\rm{F}}^{\ast}}
-\frac{1}{k_{\rm{F}}^2E_{\rm{F}}}\right),~~\Psi=\frac{2C_0k_{\rm{F}}^4}{\pi^2}\left(\frac{1}{p_{\rm{F}}F_{\rm{F}}^{\ast}}-\frac{1}{k_{\rm{F}}E_{\rm{F}}^{\ast}}\right),
\end{equation}
then, \begin{equation} \frac{\partial\rho_{\rm{S}}}{\partial\rho}
=\frac{\partial
f_0}{\partial\rho}\left[g_{\sigma}\left(\frac{\rho_{\rm{S}}^{\rm{II}}}{M_0^{\ast}}+\Psi\right)
-3g_{\sigma}\left(\frac{\rho_{\rm{S}}^{\rm{I}}}{M_0^{\ast}}-\frac{\Delta_0\rho}{{E}_{\rm{F}}^{\ast}}\right)\right]
+\frac{\Delta_0M_0^{\ast}}{{E}_{\rm{F}}^{\ast}}+\frac{4\rho_{\rm{S}}^{\rm{II}}}{3\rho}+\Phi.\end{equation}
According to the equation of motion of $f_0$, we have
\begin{equation}
m_{\sigma}^2\frac{\partial f_0}{\partial\rho} =g_{\sigma}
\frac{\partial\rho_{\rm{S}}}{\partial\rho}
-2b_{\sigma}Mg_{\sigma}^3f_0\frac{\partial
f_0}{\partial\rho}-3c_{\sigma}g_{\sigma}^4f_0^2\frac{\partial
f_0}{\partial\rho}.
\end{equation}
Thus we obtain the following expression for
$\partial f_0/\partial\rho$
\begin{equation}\label{f0withrho}
\frac{\partial
f_0}{\partial\rho}=\frac{g_{\sigma}}{R_{\sigma}}\left(\frac{\Delta_0M_0^{\ast}}{{E}_{\rm{F}}^{\ast}}+\frac{4\rho_{\rm{S}}^{\rm{II}}}{3\rho}
+\Phi\right),
\end{equation}
with
\begin{equation}
R_{\sigma}=m_{\sigma}^2+3g_{\sigma}^2\left(\frac{\rho_{\rm{S}}^{\rm{I}}}{M_0^{\ast}}
-\frac{\Delta_0\rho}{{E}_{\rm{F}}^{\ast}}\right)+2b_{\sigma}Mg_{\sigma}^3f_0+3c_{\sigma}g_{\sigma}^4f_0^2
-g_{\sigma}^2\left(\frac{\rho_{\rm{S}}^{\rm{II}}}{M_0^{\ast}}+\Psi\right).
\end{equation}

By taking derivatives term by term in the kinetic symmetry energy of (\ref{EsymkinHMT}), we obtain the following
kinetic slope parameter
\begin{align}
L^{\rm{kin}}(\rho)=&\left[\frac{k_{\rm{F}}^2({E}_{\rm{F}}^{\ast,2}+M_0^{\ast,2})}{6{E}_{\rm{F}}^{\ast,3}}
+\frac{g_{\sigma}k_{\rm{F}}^2M_0^{\ast}\rho}{2E_{\rm{F}}^{\ast,3}}\frac{\partial
f_0}{\partial\rho}\right]{}\left[1-3C_0\left(1-\frac{1}{\phi_0}\right)\right]\notag\\
&-\frac{9\rho}{E_{\rm{F}}^{\ast}}\left(\frac{\pi^2}{2k_{\rm{F}}}-g_{\sigma}M_0^{\ast}\frac{\partial
f_0}{\partial\rho}\right){}
C_0\left[C_1\left(1-\frac{1}{\phi_0}\right)+\frac{\phi_1}{\phi_0}\right]\notag\\
&-\frac{9C_0\phi_1(C_1-\phi_1)}{4\pi^2k_{\rm{F}}\phi_0E_{\rm{F}}^{\ast}}{}\Bigg[
\sqrt{1+\theta^2}{}\rm{arcsinh}\,\theta{}\left(\frac{3M_0^{\ast,5}\pi^2}{2k_{\rm{F}}^2}+4g_{\sigma}k_{\rm{F}}M_0^{\ast,4}\frac{\partial
f_0}{\partial\rho}\right)\notag\\
&\hspace*{2.cm}+\frac{\pi^2}{2k_{\rm{F}}^2}{}\left(2k_{\rm{F}}^5
-k_{\rm{F}}^3M_0^{\ast,2}-3k_{\rm{F}}M_0^{\ast,4}\right)
-4g_{\sigma}k_{\rm{F}}^2M_0^{\ast} E_{\rm{F}}^{\ast,2}\frac{\partial
f_0}{\partial\rho} \Bigg]\notag\\
&+\frac{2k_{\rm{F}}C_0(6C_1+1)}{3}{}\left[\rm{arcsinh}\,(\phi_0\theta)-\sqrt{1+\frac{1}{\phi_0^2\theta^2}}
-\rm{arcsinh}\,\theta+\sqrt{1+\frac{1}{\theta^2}}\right]\notag\\
&+{2k_{\rm{F}}\rho
C_0(6C_1+1)}{}\left(\frac{M_0^{\ast}\pi^2}{2k_{\rm{F}}^2}+g_{\sigma}k_{\rm{F}}\frac{\partial
f_0}{\partial\rho}\right)\notag\\
&\hspace*{2.cm}\times\left(\frac{\phi_0}{F_{\rm{F}}^{\ast}M_0^{\ast}}
+\frac{M_0^{\ast}}{\phi_0^2k_{\rm{F}}^3\sqrt{1+\frac{\displaystyle1}{\displaystyle\phi_0^2\theta^2}}}
-\frac{1}{M_0^{\ast,2}\sqrt{1+\theta^2}}-\frac{1}{M_0^{\ast}k_{\rm{F}}\theta^2\sqrt{1+\frac{\displaystyle1}{\displaystyle\theta^2}}}\right).
\notag\\
&+\frac{3k_{\rm{F}}C_0}{2}{}\left[\frac{(1+3\phi_1)^2}{9}\left(\frac{\phi_0k_{\rm{F}}}{F_{\rm{F}}^{\ast}}
-\frac{2F_{\rm{F}}^{\ast}}{\phi_0k_{\rm{F}}}\right)+\frac{2F_{\rm{F}}^{\ast}(3\phi_1-1)}{9\phi_0k_{\rm{F}}}
-\frac{1}{9}\frac{k_{\rm{F}}}{E_{\rm{F}}^{\ast}}+\frac{4E_{\rm{F}}^{\ast}}{9k_{\rm{F}}}\right]\notag\\
&+\frac{9k_{\rm{F}}\rho
C_0}{2}{}\left(\frac{M_0^{\ast}\pi^2}{2k_{\rm{F}}^2}+g_{\sigma}k_{\rm{F}}\frac{\partial
f_0}{\partial\rho}\right){}\Bigg[\frac{(1+3\phi_1)^2}{9}\left(\frac{\phi_0}{F_{\rm{F}}^{\ast}M_0^{\ast}}
-\frac{\phi_0^3\theta^2}{M_0^{\ast,2}(1+\phi_0^2\theta^2)^{3/2}}+\frac{2}{\phi_0k_{\rm{F}}^2\sqrt{1+\phi_0^2\theta^2}}\right)\notag\\
&\hspace*{2cm}-\frac{2(3\phi_1-1)}{9\phi_0k_{\rm{F}}^2\sqrt{1+\phi_0^2\theta^2}}-\frac{M_0^{\ast}}{9E_{\rm{F}}^3}-\frac{4}{9k_{\rm{F}}^2\sqrt{1+\theta^2}}
\Bigg]\notag\\
&+{\rho C_0(4+3C_1)}{}\Bigg[
\frac{\pi^2}{2k_{\rm{F}}^2}{}\left[\frac{(1+3\phi_1)\phi_0k_{\rm{F}}}{M_0^{\ast}\sqrt{1+\phi_0^2\theta^2}}
-\frac{k_{\rm{F}}}{M_0^{\ast}\sqrt{1+\theta^2}}\right]\notag\\
&\hspace*{2.cm}-g_{\sigma}\frac{\partial
f}{\partial\rho}_0{}\left[\frac{(1+3\phi_1)\sqrt{1+\phi_0^2\theta^2}}{\phi_0}
-\frac{(1+3\phi_1)\phi_0k_{\rm{F}}^2}{M_0^{\ast,2}\sqrt{1+\phi_0^2\theta^2}}
-\sqrt{1+\theta^2}+\frac{k_{\rm{F}}^2}{M_0^{\ast,2}
\sqrt{1+\theta^2}}\right]\Bigg], \label{Lkin}
\end{align}where $\partial f_0/\partial\rho$ is given by (\ref{f0withrho}) and
\begin{align}
\rho_{\rm{S}}^{\rm{I}}&=\frac{\Delta_0M_0^{\ast,3}}{\pi^2}
\left(\theta\sqrt{1+\theta^2}-\rm{arcsinh}\,\theta\right),~~
\rho_{\rm{S}}^{\rm{II}}
=\frac{2C_0k_{\rm{F}}^4}{\pi^2M_0^{\ast}}\left(\sqrt{1+\frac{1}{\theta^2}}-\sqrt{1+\frac{1}{\phi_0^2\theta^2}}\right)
,\label{SNMRHOS}
\end{align}
with $\theta=k_{\rm{F}}/M_0^{\ast}$.
The potential part of the slope parameter of the symmetry energy is
given by
\begin{equation}
L^{\rm{pot}}(\rho)=3\rho\frac{\partial
E_{\rm{sym}}^{\rm{pot}}(\rho)}{\partial\rho}=\frac{3g_{\rho}^2\rho}{2Q_{\rho}}
-\frac{3g_{\omega}^3g_{\rho}^4\Lambda_{\rm{V}}w_0\rho^2}{Q_{\omega}Q_{\rho}^2}\label{Lpot}
\end{equation}
with
\begin{equation}Q_{\omega}=m_{\omega}^2+3c_{\omega}g_{\omega}^4w_0\end{equation} and
$Q_{\rho}$ given by (\ref{Esympot}). The total slope parameter of
the symmetry energy is given by
\begin{equation}\label{TOTL}
L(\rho)=L^{\rm{kin}}(\rho)+L^{\rm{pot}}(\rho).
\end{equation}

\end{widetext}

\section{The Derivation of $K_0(\rho)$}\label{App3}

The incompressibility coefficient $K_0\equiv K_0(\rho)$ can be
obtained through
\begin{align}
K_0(\rho)=&9\rho^2\frac{\partial^2E_0}{\partial\rho^2}
=9\frac{\partial P_0}{\partial\rho} -\frac{18P_0}{\rho},
\end{align}
where $P_0(\rho)$ is the pressure of SNM. At normal density, the
pressure of SNM is zero, thus only the first term of the last
expression is relevant for our aim. So we should first calculate the
pressure $P_0$ as a function of density. Before moving on, we first
prove the following relation
\begin{equation}\label{HVH}
P_0(\rho)=\rho^2\frac{\partial
E_0(\rho)}{\partial\rho}=\rho^2\frac{\partial(\varepsilon_0/\rho)}{\partial\rho}
\end{equation}
by calculating the quantities on both sides and then comparing them.
\begin{widetext}

The EOS of SNM is obtained from the energy density
\begin{align}
\varepsilon_0=&2\varepsilon^{\rm{kin}}_{\wp,0}+\frac{1}{2}m_{\sigma}^2f_0^2+\frac{1}{2}m_{\omega}^2w_0^2
+\frac{1}{3}b_{\sigma}Mg_{\sigma}^3f_0^3
+\frac{1}{4}c_{\sigma}g_{\sigma}^4f_0^4+\frac{3}{4}c_{\omega}g_{\omega}^4w_0^4,\label{eps0rho}\end{align}
with
\begin{align}
\varepsilon_{\wp,0}^{\rm{kin}}
=&\frac{2}{(2\pi)^3}\int_0^{\phi_0k_{\rm{F}}}n_{\v{k}}^0\sqrt{\v{k}^2+M_0^{\ast,2}}\d\v{k}
=\frac{1}{\pi^2}\left[\Delta_0\int_0^{k_{\rm{F}}}k^2\d
k\sqrt{k^2+M_0^{\ast,2}}+{C_0k_{\rm{F}}^4}\int_{k_{\rm{F}}}^{\phi_0k_{\rm{F}}}\frac{1}{k^2}\sqrt{k^2+M_0^{\ast,2}}\d
k\right],\end{align} here $\wp$ is just a symbol reminding us that
$\varepsilon_{\wp,0}^{\rm{kin,I}}+\varepsilon_{\wp,0}^{\rm{kin,II}}=\varepsilon_{\wp,0}^{\rm{kin}}=2^{-1}\varepsilon_0^{\rm{kin}}$,
where $\varepsilon_0^{\rm{kin}}$ is the total kinetic energy density
(including n and p). Similarly, the pressure is\begin{equation}
P_0=2P_{\wp,0}^{\rm{kin}}-\frac{1}{2}m_{\sigma}^2f_0^2+\frac{1}{2}m_{\omega}^2w_0^2
-\frac{1}{3}b_{\sigma}Mg_{\sigma}^3f_0^3-\frac{1}{4}c_{\sigma}g_{\sigma}^4f_0^4+\frac{1}{4}c_{\omega}g_{\omega}^4w_0^4,\end{equation}
with
\begin{equation}
P_{\wp,0}^{\rm{kin}}=\frac{1}{3\pi^2}\left[\Delta_0\int_0^{k_{\rm{F}}}\d
k\frac{k^4}{\sqrt{k^2+M_0^{\ast,2}}}+{C_0k_{\rm{F}}^4}\int_{k_{\rm{F}}}^{\phi_0k_{\rm{F}}}\d
k\frac{1}{\sqrt{k^2+M_0^{\ast,2}}}\right].\end{equation}

For the $\omega$ field, we have ${\partial
w_0}/{\partial\rho}={g_{\omega}}/{Q_{\omega}} $. The $\omega$ part
in the energy density has the following relation
\begin{align}
\rho^2\frac{\partial}{\partial\rho}\left(\frac{1}{2}m_{\omega}^2w_0^2+\frac{3}{4}c_{\omega}g_{\omega}^4w_0^4\right)
=&\frac{1}{2}m_{\omega}^2w_0^2+\frac{1}{4}c_{\omega}g_{\omega}^4w_0^4,
\end{align}
which are just the corresponding terms of the pressure.

The proof for $\sigma$ field is somewhat more complicated. The first
part of the kinetic energy density is given by
\begin{align}
\varepsilon_{\wp,0}^{\rm{kin,I}}=&\frac{\Delta_0M_0^{\ast,4}}{\pi^2}\int_0^{k_{\rm{F}}/M_0^{\ast}}\zeta^2\sqrt{1+\zeta^2}\d\zeta
=\frac{\Delta_0M_0^{\ast,4}}{\pi^2}\left[\frac{1}{4}\theta(1+\theta^2)^{3/2}-\frac{1}{8}\theta\sqrt{1+\theta^2}-\frac{1}{8}\rm{arcsinh}\,\theta\right],
\label{eps0rho1}\end{align} thus
\begin{align}
\frac{\partial\varepsilon_{\wp,0}^{\rm{kin,I}}}{\partial\rho}
=&\frac{\Delta_0}{\pi^2}\cdot4M_0^{\ast,3}\frac{\partial
M_0^{\ast}}{\partial\rho}\int_0^{k_{\rm{F}}/M_0^{\ast}}\zeta^2\sqrt{1+\zeta^2}\d\zeta
+\frac{\Delta_0M_0^{\ast,4}}{\pi^2}\frac{\partial}{\partial\rho}\int_0^{k_{\rm{F}}/M_0^{\ast}}\zeta^2\sqrt{1+\zeta^2}\d\zeta\notag\\
=&\frac{4\varepsilon_{\wp,0}^{\rm{kin,I}}}{M_0^{\ast}}\frac{\partial
M_0^{\ast}}{\partial\rho}+\frac{\Delta_0M_0^{\ast,4}}{\pi^2}\cdot\frac{k_{\rm{F}}^2}{M_0^{\ast,3}}\frac{E_{\rm{F}}^{\ast}}{M_0^{\ast,2}}
\left(M_0^{\ast}\frac{\partial
k_{\rm{F}}}{\partial\rho}-k_{\rm{F}}\frac{\partial
M_0^{\ast}}{\partial\rho}\right)\notag\\
=&\frac{4\varepsilon_{\wp,0}^{\rm{kin,I}}}{M_0^{\ast}}\frac{\partial
M_0^{\ast}}{\partial\rho}+\frac{\Delta_0k_{\rm{F}}^2E_{\rm{F}}^{\ast}}{\pi^2M_0^{\ast}}\left(M_0^{\ast}\frac{\partial
k_{\rm{F}}}{\partial\rho}-k_{\rm{F}}\frac{\partial
M_0^{\ast}}{\partial\rho}\right)\notag\\
=&\left(\frac{4\varepsilon_{\wp,0}^{\rm{kin,I}}}{M_0^{\ast}}\frac{\partial
M_0^{\ast}}{\partial\rho}-\frac{\Delta_0k_{\rm{F}}^3E_{\rm{F}}^{\ast}}{\pi^2M_0^{\ast}}\right)\frac{\partial
M_0^{\ast}}{\partial\rho}+\frac{\Delta_0k_{\rm{F}}^2E_{\rm{F}}^{\ast}}{\pi^2}\frac{\partial
k_{\rm{F}}}{\partial\rho}\notag\\
=&\left(\frac{4\varepsilon_{\wp,0}^{\rm{kin,I}}}{M_0^{\ast}}\frac{\partial
M_0^{\ast}}{\partial\rho}-\frac{\Delta_0k_{\rm{F}}^3E_{\rm{F}}^{\ast}}{\pi^2M_0^{\ast}}\right)\frac{\partial
M_0^{\ast}}{\partial\rho}+\frac{\Delta_0E_{\rm{F}}^{\ast}}{2}.
\end{align}
Similarly,
\begin{align}
\varepsilon_{\wp,0}^{\rm{kin,II}}
=&\frac{C_0k_{\rm{F}}^4}{\pi^2}\int_{k_{\rm{F}}/M_0^{\ast}}^{p_{\rm{F}}/M_0^{\ast}}
\frac{\sqrt{1+\zeta^2}}{\zeta^2}\d\zeta
=\frac{C_0k_{\rm{F}}^4}{\pi^2}\left[
\rm{arcsinh}(\phi_0\theta)-\rm{arcsinh}\,\theta-\sqrt{1+\frac{1}{\phi_0^2\theta^2}}+\sqrt{1+\frac{1}{\theta^2}}\right]
,\label{eps0rho2}\end{align} and
\begin{align}
\frac{\partial\varepsilon_{\wp,0}^{\rm{kin,II}}}{\partial\rho}
=&\frac{C_0}{\pi^2}\cdot4k_{\rm{F}}^3\frac{\partial
k_{\rm{F}}}{\partial\rho}\int_{k_{\rm{F}}/M_0^{\ast}}^{p_{\rm{F}}/M_0^{\ast}}
\frac{\sqrt{1+\zeta^2}}{\zeta^2}\d\zeta+\frac{C_0k_{\rm{F}}^4}{\pi^2}\frac{\partial}{\partial\rho}
\int_{k_{\rm{F}}/M_0^{\ast}}^{p_{\rm{F}}/M_0^{\ast}}
\frac{\sqrt{1+\zeta^2}}{\zeta^2}\d\zeta\notag\\
=&\left[\frac{4\varepsilon_{\wp,0}^{\rm{kin,II}}}{k_{\rm{F}}}+\frac{C_0k_{\rm{F}}^4}{\pi^2}\left(\frac{\phi_0F_{\rm{F}}^{\ast}}{p_{\rm{F}}^2}
-\frac{E_{\rm{F}}^{\ast}}{k_{\rm{F}}^2}\right)\right]\frac{\partial
k_{\rm{F}}}{\partial\rho}
-\frac{C_0k_{\rm{F}}^4}{\pi^2}\frac{1}{M_0^{\ast}}\left(
\frac{F_{\rm{F}}^{\ast}}{p_{\rm{F}}}-\frac{E_{\rm{F}}^{\ast}}{k_{\rm{F}}}\right)\frac{\partial
M_0^{\ast}}{\partial\rho}.
\end{align}
Putting into the expression for $\varepsilon_{\wp,0}^{\rm{kin,II}}$,
we have
\begin{align}
\frac{\partial\varepsilon_{\wp,0}^{\rm{kin,II}}}{\partial\rho}=&
\frac{4}{k_{\rm{F}}}\frac{\partial
k_{\rm{F}}}{\partial\rho}\left[\rm{arcsinh}(\phi_0\theta)-\rm{arcsinh}\,\theta-\sqrt{1+\frac{1}{\phi_0^2\theta^2}}+\sqrt{1+\frac{1}{\theta^2}}\right]\notag\\
&+\frac{C_0k_{\rm{F}}^4}{\pi^2}\left(\frac{\phi_0F_{\rm{F}}^{\ast}}{p_{\rm{F}}^2}
-\frac{E_{\rm{F}}^{\ast}}{k_{\rm{F}}^2}\right)\frac{\partial
k_{\rm{F}}}{\partial\rho}-\frac{C_0k_{\rm{F}}^4}{\pi^2}\frac{1}{M_0^{\ast}}\left(
\frac{F_{\rm{F}}^{\ast}}{p_{\rm{F}}}-\frac{E_{\rm{F}}^{\ast}}{k_{\rm{F}}}\right)\frac{\partial
M_0^{\ast}}{\partial\rho}.
\end{align}

Then we have (only the $f_0$ field is considered here)
\begin{align}
\rho^2\frac{\partial}{\partial\rho}\left(\frac{\varepsilon_0^{f_0}}{\rho}\right)=
&\rho\frac{\partial\varepsilon_0^{f_0}}{\partial\rho} -\varepsilon_0^{f_0}\notag\\
=&-\frac{1}{2}m_{\sigma}^2f_0^2-\frac{1}{3}b_{\sigma}Mg_{\sigma}^3f_0^3-\frac{1}{4}c_{\sigma}g_{\sigma}^4f_0^4
+\rho\frac{\partial
f_0}{\partial\rho}\left(m_{\sigma}^2f_0+b_{\sigma}Mg_{\sigma}^3f_0^2+c_{\sigma}g_{\sigma}^4f_0^3\right)\notag\\
&-2\Bigg[\frac{\Delta_0M_0^{\ast,4}}{\pi^2}\left[\frac{1}{4}\theta(1+\theta^2)^{3/2}-\frac{1}{8}\theta\sqrt{1+\theta^2}-\frac{1}{8}\rm{arcsinh}\,\theta\right]\notag\\
&\hspace*{0.cm}+\frac{C_0k_{\rm{F}}^4}{\pi^2}\left[
\rm{arcsinh}(\phi_0\theta)-\rm{arcsinh}\,\theta-\sqrt{1+\frac{1}{\phi_0^2\theta^2}}+\sqrt{1+\frac{1}{\theta^2}}\right]\Bigg]\notag\\
&+2\rho\Bigg[-g_{\sigma}\frac{\partial
f_0}{\partial\rho}\left(\frac{4\varepsilon_{\wp,0}^{\rm{kin,I}}}{M_0^{\ast}}
-\frac{\Delta_0k_{\rm{F}}^3E_{\rm{F}}^{\ast}}{\pi^2M_0^{\ast}}\right)
+\frac{\Delta_0E_{\rm{F}}^{\ast}}{2}\notag\notag\\
&\hspace*{0.cm}
+\left[\frac{4\varepsilon_{\wp,0}^{\rm{kin,II}}}{k_{\rm{F}}}+\frac{C_0k_{\rm{F}}^4}{\pi^2}\left(\frac{\phi_0F_{\rm{F}}^{\ast}}{p_{\rm{F}}^2}
-\frac{E_{\rm{F}}^{\ast}}{k_{\rm{F}}^2}\right)\right]\frac{\partial
k_{\rm{F}}}{\partial\rho}\notag
+\frac{g_{\sigma}}{M_0^{\ast}}\frac{C_0k_{\rm{F}}^4}{\pi^2}\left(
\frac{F_{\rm{F}}^{\ast}}{p_{\rm{F}}}-\frac{E_{\rm{F}}^{\ast}}{k_{\rm{F}}}\right)\frac{\partial
f_0}{\partial\rho}\Bigg].
\end{align}

The term proportional to $\partial f_0/\partial\rho$ in
$\rho^2\partial(\varepsilon_0/\rho)/\partial\rho$ is
\begin{align}
\Pi_0=&\rho\left(m_{\sigma}^2f_0+b_{\sigma}Mg_{\sigma}^3f_0^2+c_{\sigma}g_{\sigma}^4f_0^3\right)
-2g_{\sigma}\rho\left(\frac{4\varepsilon_{\wp,0}^{\rm{kin,I}}}{M_0^{\ast}}
-\frac{\Delta_0k_{\rm{F}}^3E_{\rm{F}}^{\ast}}{\pi^2M_0^{\ast}}\right)
+\frac{2g_{\sigma}\rho}{M_0^{\ast}}\frac{C_0k_{\rm{F}}^4}{\pi^2}\left(
\frac{F_{\rm{F}}^{\ast}}{p_{\rm{F}}}
-\frac{E_{\rm{F}}^{\ast}}{k_{\rm{F}}}\right),\end{align} where
\begin{equation}
\varepsilon_{\wp,0}^{\rm{kin,I}}=\frac{\Delta_0k_{\rm{F}}^3E_{\rm{F}}^{\ast}}{4\pi^2}+\frac{M_0^{\ast}\rho_{\rm{S},\wp,0}^{\rm{I}}}{4},~~\frac{4\varepsilon_{\wp,0}^{\rm{kin,I}}}{M_0^{\ast}}
-\frac{\Delta_0k_{\rm{F}}^3E_{\rm{F}}^{\ast}}{\pi^2M_0^{\ast}}=\rho_{\rm{S},\wp,0}^{\rm{I}},
\end{equation}
Similarly, we have
\begin{equation}
\rho_{\rm{S},\wp,0}^{\rm{II}}=-\frac{C_0k_{\rm{F}}^4}{\pi^2M_0^{\ast}}\left(\frac{F_{\rm{F}}^{\ast}}{p_{\rm{F}}}-
\frac{E_{\rm{F}}^{\ast}}{k_{\rm{F}}}\right),\end{equation}thus
\begin{align} \Pi_0=&
\rho\left(m_{\sigma}^2f_0+b_{\sigma}Mg_{\sigma}^3f_0^2+c_{\sigma}g_{\sigma}^4f_0^3\right)
-2g_{\sigma}\rho_{\rm{S},\wp,0}^{\rm{I}}-2g_{\sigma}\rho_{\rm{S},\wp,0}^{\rm{II}}
=\rho\left[\left(m_{\sigma}^2f_0+b_{\sigma}Mg_{\sigma}^3f_0^2+c_{\sigma}g_{\sigma}^4f_0^3\right)-g_{\sigma}\rho_{\rm{S}}\right],\end{align}
where
$\rho_{\rm{S}}=2(\rho_{\rm{S},\wp,0}^{\rm{I}}+\rho_{\rm{S},\wp,0}^{\rm{II}})$
is the total scalar density, the above equation equals to zero according
to the equation of motion of $f_0$. Thus
\begin{align}
\rho^2\frac{\partial}{\partial\rho}\left(\frac{\varepsilon_0^{f_0}}{\rho}\right)
=&-\frac{1}{2}m_{\sigma}^2f_0^2-\frac{1}{3}b_{\sigma}Mg_{\sigma}^3f_0^3-\frac{1}{4}c_{\sigma}g_{\sigma}^4f_0^4\notag\\
&-2\Bigg[\frac{\Delta_0M_0^{\ast,4}}{\pi^2}\left[\frac{1}{4}\theta(1+\theta^2)^{3/2}-\frac{1}{8}\theta\sqrt{1+\theta^2}-\frac{1}{8}\rm{arcsinh}\,\theta\right]\notag\\
&\hspace*{0.cm}+\frac{C_0k_{\rm{F}}^4}{\pi^2}\left[
\rm{arcsinh}(\phi_0\theta)-\rm{arcsinh}\,\theta-\sqrt{1+\frac{1}{\phi_0^2\theta^2}}+\sqrt{1+\frac{1}{\theta^2}}\right]\Bigg]\notag\\
&+2\rho\left[\frac{\Delta_0E_{\rm{F}}^{\ast}}{2}
+\left[\frac{4\varepsilon_{\wp,0}^{\rm{kin,II}}}{k_{\rm{F}}}+\frac{C_0k_{\rm{F}}^4}{\pi^2}\left(\frac{\phi_0F_{\rm{F}}^{\ast}}{p_{\rm{F}}^2}
-\frac{E_{\rm{F}}^{\ast}}{k_{\rm{F}}^2}\right)\right]\frac{\partial
k_{\rm{F}}}{\partial\rho}\right]\label{e0line333}.
\end{align}

The expression for pressure is easy to obtain,
\begin{align}
P_{\wp,0}^{\rm{kin,I}}
=&\frac{\Delta_0M_0^{\ast,4}}{3\pi^2}\int_0^{k_{\rm{F}}/M_0^{\ast}}\frac{\zeta^4\d\zeta}{\sqrt{1+\zeta^2}}
=\frac{\Delta_0M_0^{\ast,4}}{3\pi^2}\left[\frac{1}{4}\theta^3\sqrt{1+\theta^2}
-\frac{3}{8}\theta\sqrt{1+\theta^2}+\frac{3}{8}\rm{arcsinh}\,\theta\right],\\
P_{\wp,0}^{\rm{kin,II}}
=&\frac{C_0k_{\rm{F}}^4}{3\pi^2}\int_{k_{\rm{F}}/M_0^{\ast}}^{p_{\rm{F}}/M_0^{\ast}}
\frac{\d\zeta}{\sqrt{1+\zeta^2}}=\frac{C_0k_{\rm{F}}^4}{3\pi^2}
\left[\rm{arcsinh}(\phi_0\theta)-\rm{arcsinh}\,\theta\right],
\end{align}
thus
\begin{align}
P_0^{f_0}=&-\frac{1}{2}m_{\sigma}^2f_0^2-\frac{1}{3}b_{\sigma}Mg_{\sigma}^3f_0^3-\frac{1}{4}c_{\sigma}g_{\sigma}^4f_0^4\notag\\
&+\frac{2\Delta_0M_0^{\ast,4}}{3\pi^2}\left[\frac{1}{4}\theta^3\sqrt{1+\theta^2}
-\frac{3}{8}\theta\sqrt{1+\theta^2}+\frac{3}{8}\rm{arcsinh}\,\theta\right]\notag\\
&+\frac{2C_0k_{\rm{F}}^4}{3\pi^2}
\left[\rm{arcsinh}(\phi_0\theta)-\rm{arcsinh}\,\theta\right]\label{p0line333}.
\end{align}
It is obvious that the first line of (\ref{e0line333}) and that of
(\ref{p0line333}) are the same. Terms proportional to $\Delta_0$ in
(\ref{e0line333}) are
\begin{align}
&\Delta_0\rho
E_{\rm{F}}^{\ast}-\frac{2\Delta_0M_0^{\ast,4}}{\pi^2}\left[\frac{1}{4}\theta(1+\theta^2)^{3/2}
-\frac{1}{8}\theta\sqrt{1+\theta^2}-\frac{1}{8}\rm{arcsinh}\,\theta\right]\notag\\
=&\frac{2\Delta_0k_{\rm{F}}^3E_{\rm{F}}^{\ast}}{3\pi^2}-\frac{2\Delta_0M_0^{\ast,4}}{\pi^2}\left[\frac{1}{4}\theta(1+\theta^2)^{3/2}
-\frac{1}{8}\theta\sqrt{1+\theta^2}-\frac{1}{8}\rm{arcsinh}\,\theta\right]\notag\\
=&\frac{2\Delta_0M_0^{\ast,4}}{3\pi^2}\left[\frac{1}{4}\theta^3\sqrt{1+\theta^2}-\frac{3}{8}\theta\sqrt{1+\theta^2}
+\frac{3}{8}\rm{arcsinh}\,\theta\right],
\end{align}
this is the corresponding term in the pressure proportional to
$\Delta_0$. Similarly, the remaining terms in (\ref{e0line333}) are
\begin{align}\label{d1}
&2\rho\frac{\partial
k_{\rm{F}}}{\partial\rho}\left[\frac{4\varepsilon_{\wp,0}^{\rm{kin,II}}}{k_{\rm{F}}}+\frac{C_0k_{\rm{F}}^4}{\pi^2}\left(\frac{\phi_0F_{\rm{F}}^{\ast}}{p_{\rm{F}}^2}
-\frac{E_{\rm{F}}^{\ast}}{k_{\rm{F}}^2}\right)\right]\notag\\
=&\frac{2C_0k_{\rm{F}}^4}{3\pi^2}\left[4\left(
\rm{arcsinh}(\phi_0\theta)-\rm{arcsinh}\,\theta-\sqrt{1+\frac{1}{\phi_0^2\theta^2}}+\sqrt{1+\frac{1}{\theta^2}}\right)\right.
+\left.\left(\sqrt{1+\frac{1}{\phi_0^2\theta^2}}-\sqrt{1+\frac{1}{\theta^2}}\right)
\right].
\end{align}
and (i.e., the third line of (\ref{e0line333}))
\begin{equation}\label{d2}
-\frac{2C_0k_{\rm{F}}^4}{3\pi^2}\cdot3\left(
\rm{arcsinh}(\phi_0\theta)-\rm{arcsinh}\,\theta-\sqrt{1+\frac{1}{\phi_0^2\theta^2}}+\sqrt{1+\frac{1}{\theta^2}}\right),\end{equation}
combining (\ref{d1}) and (\ref{d2}), we obtain the following
expression proportional to $C_0$ in (\ref{e0line333}),
\begin{equation}
\frac{2C_0k_{\rm{F}}^4}{3\pi^2}[\rm{arcsinh}(\phi_0\theta)-\rm{arcsinh}\,\theta],\end{equation}
which is exactly the same as the last line of (\ref{p0line333}).
Thus we have proved the relation (\ref{HVH}).

The expression for pressure including $w_0$ field then reads
\begin{align}
P_0=&-\frac{1}{2}m_{\sigma}^2f_0^2-\frac{1}{3}b_{\sigma}Mg_{\sigma}^3f_0^3-\frac{1}{4}c_{\sigma}g_{\sigma}^4f_0^4
+\frac{1}{2}m_{\omega}^2w_0^2+\frac{1}{4}c_{\omega}g_{\omega}^4w_0^4\notag\\
&+\frac{2\Delta_0M_0^{\ast,4}}{3\pi^2}\left[\frac{1}{4}\theta^3\sqrt{1+\theta^2}
-\frac{3}{8}\theta\sqrt{1+\theta^2}+\frac{3}{8}\rm{arcsinh}\,\theta\right]\notag\\
&+\frac{2C_0k_{\rm{F}}^4}{3\pi^2}
\left[\rm{arcsinh}(\phi_0\theta)-\rm{arcsinh}\,\theta\right].\label{p0rho}
\end{align}
The contribution to $K_0$ from $w_0$ field is
\begin{equation}
9\frac{\partial}{\partial\rho}\left(\frac{1}{2}m_{\omega}^2w_0^2+\frac{1}{4}c_{\omega}g_{\omega}^4w_0^4\right)
=\frac{9\rho g_{\omega}^2}{Q_{\omega}}.\end{equation} For the $f_0$
field, we have
\begin{align}
\frac{\partial P_0^{f_0}}{\partial\rho} =&-f_0\frac{\partial
f_0}{\partial\rho}(m_{\sigma}^2+b_{\sigma}Mg_{\sigma}^3f_0+c_{\sigma}g_{\sigma}^4f_0^2)
+\frac{2\Delta_0}{3\pi^2}\left[W(\theta)\frac{\partial
M_0^{\ast,4}}{\partial\rho}+M_0^{\ast,4}\frac{\partial
W(\theta)}{\partial\rho}\right]+\frac{2C_0}{3\pi^2}\left[V(\theta)\frac{\partial
k_{\rm{F}}^{4}}{\partial\rho}+k_{\rm{F}}^4\frac{\partial
V(\theta)}{\partial\rho}\right]\notag\\
=&-\frac{\partial f_0}{\partial\rho}
\left[f_0(m_{\sigma}^2+b_{\sigma}Mg_{\sigma}^3f_0+c_{\sigma}g_{\sigma}^4f_0^2)
+\frac{8g_{\sigma}\Delta_0M_0^{\ast,3}W(\theta)}{3\pi^2}\right]
+\frac{4C_0k_{\rm{F}}V(\theta)}{3}\notag\\
&+\left[\frac{2\Delta_0k_{\rm{F}}^4}{3\pi^2M_0^{\ast}E_{\rm{F}}^{\ast}}
+\frac{2C_0k_{\rm{F}}^4}{3\pi^2M_0^{\ast}}\left(\frac{\phi_0}{F_{\rm{F}}^{\ast}}-\frac{1}{E_{\rm{F}}^{\ast}}\right)\right]\left(M_0^{\ast}\frac{\partial
k_{\rm{F}}}{\partial\rho}-k_{\rm{F}}\frac{\partial
M_0^{\ast}}{\partial\rho}\right),
\end{align}
where
\begin{align}
W(\theta)=&\frac{1}{4}\theta^3\sqrt{1+\theta^2}
-\frac{3}{8}\theta\sqrt{1+\theta^2}+\frac{3}{8}\rm{arcsinh}\,\theta,~~
V(\theta)=\rm{arcsinh}(\phi_0\theta)-\rm{arcsinh}\,\theta.\end{align}
The corresponding contribution to the $K_0$ of $f_0$ field is
\begin{align}
&-9\frac{\partial f_0}{\partial\rho}
\left[f_0(m_{\sigma}^2+b_{\sigma}Mg_{\sigma}^3f_0+c_{\sigma}g_{\sigma}^4f_0^2)
+\frac{8g_{\sigma}\Delta_0M_0^{\ast,3}W(\theta)}{3\pi^2}\right]
+12C_0k_{\rm{F}}V(\theta)\notag\\
&+9\left[\frac{2\Delta_0k_{\rm{F}}^4}{3\pi^2M_0^{\ast}E_{\rm{F}}^{\ast}}
+\frac{2C_0k_{\rm{F}}^4}{3\pi^2M_0^{\ast}}\left(\frac{\phi_0}{F_{\rm{F}}^{\ast}}-\frac{1}{E_{\rm{F}}^{\ast}}\right)\right]\left(M_0^{\ast}\frac{\partial
k_{\rm{F}}}{\partial\rho}-k_{\rm{F}}\frac{\partial
M_0^{\ast}}{\partial\rho}\right).
\end{align}
Combining the above results, we finally obtain the expression for
$K_0(\rho)$ as
\begin{align}
K_0(\rho)=&-9\frac{\partial f_0}{\partial\rho}
\left[f_0(m_{\sigma}^2+b_{\sigma}Mg_{\sigma}^3f_0+c_{\sigma}g_{\sigma}^4f_0^2)
+\frac{8g_{\sigma}\Delta_0M_0^{\ast,3}W(\theta)}{3\pi^2}\right]\notag\\
&+9\left[\frac{2\Delta_0k_{\rm{F}}^4}{3\pi^2M_0^{\ast}E_{\rm{F}}^{\ast}}
+\frac{2C_0k_{\rm{F}}^4}{3\pi^2M_0^{\ast}}\left(\frac{\phi_0}{F_{\rm{F}}^{\ast}}-\frac{1}{E_{\rm{F}}^{\ast}}\right)\right]\left(\frac{M_0^{\ast}\pi^2
}{2k_{\rm{F}}^2}+g_{\sigma}k_{\rm{F}}\frac{\partial
f_0}{\partial\rho}\right)\notag\\
&+12C_0k_{\rm{F}}V(\theta)+\frac{9\rho
g_{\omega}^2}{Q_{\omega}}-\frac{18P_0}{\rho},\label{K0rho}
\end{align}
with $\partial f_0/\partial\rho$ given by (\ref{f0withrho}) and
$P_0$ by (\ref{p0rho}).
\end{widetext}

\end{document}